 \documentstyle[11pt,aaspp4]{article}

 \newcommand{\um}{\mbox{$\mu{\rm m}$}}
 \newcommand{\us}{\mbox{$\mu{\rm s}$}}
 \newcommand{\uas}{\mbox{$\mu{\rm as}$}}

 \newcommand{\p}{\mbox{$^{\prime}$}}

\begin{document}

 \slugcomment{Accepted for publication in the Astrophysical Journal. \\
 Scheduled for Vol. 510 \#1 (Jan. 1, 1999).}

 \title{The Palomar Testbed Interferometer}
 
 \author{M. M. Colavita\altaffilmark{1},
 J. K. Wallace\altaffilmark{1},
 B. E. Hines\altaffilmark{1},
 Y. Gursel\altaffilmark{1},
 F. Malbet\altaffilmark{1,3}
 D. L. Palmer\altaffilmark{1},
 X. P. Pan\altaffilmark{2},
 M. Shao\altaffilmark{1},
 J. W. Yu\altaffilmark{1},
 A. F. Boden\altaffilmark{1},
 P. J. Dumont\altaffilmark{1},
 J. Gubler\altaffilmark{4},
 C. D. Koresko\altaffilmark{2},
 S. R. Kulkarni\altaffilmark{2},
 B. F. Lane\altaffilmark{1},
 D. W. Mobley\altaffilmark{1},
 G. T. van Belle\altaffilmark{1}}

 \altaffiltext{1}{Jet Propulsion Laboratory, California Institute of Technology, 4800 Oak Grove Dr., Pasadena, CA   91109}
 \altaffiltext{2}{California Institute of Technology,  Pasadena, CA}
 \altaffiltext{3}{Currently at Laboratoire d'Astrophysique, Observatoire de Grenoble, UMR UJF/CNRS 5571, France}
 \altaffiltext{4}{University of California, San Diego, CA}


 \begin{abstract}

The Palomar Testbed Interferometer (PTI) is a long-baseline infrared interferometer located at Palomar Observatory, California.  It was built as a testbed for interferometric techniques applicable to the Keck Interferometer. First fringes were obtained in July 1995.  PTI implements a dual-star architecture, tracking two stars simultaneously for phase referencing and narrow-angle astrometry.  The three fixed 40-cm apertures can be combined pair-wise to provide baselines to 110~m.  The interferometer actively tracks the white-light fringe using an array detector at 2.2~\um\ and active delay lines with a range of $\pm$38~m.  Laser metrology of the delay lines allows for servo control, and laser metrology of the complete optical path enables narrow-angle astrometric measurements.  The instrument is highly automated, using a multiprocessing computer system for instrument control and sequencing.

 \end{abstract} 

 \section{Introduction}
 \label{sec:intro}

The Palomar Testbed Interferometer (PTI) is a long-baseline infrared interferometer installed at Palomar Observatory, California.  It was developed by the Jet Propulsion Laboratory, California Institute of Technology, for NASA as a testbed for interferometric techniques applicable to the Keck Interferometer (\cite{Keck}).  These include high-sensitivity direct-detection interferometry in the infrared with array detectors and dual-star interferometry with phase referencing for narrow-angle astrometry.  However, the technology to implement PTI, including active fringe tracking, optical delay lines, laser metrology, and real-time control, is applicable not only to the Keck Interferometer but also to other missions such as the Space Interferometry Mission (SIM) (\cite{SIM}).  PTI also serves as a testbed for interferometric planning, operational, and data-processing and management tools applicable to both missions.

Major development of PTI began in November 1992 with the commencement of funding from NASA under its TOPS program.  The interferometer was installed at Palomar Observatory during the spring of 1995, and first fringes were obtained in July 1995.  Subsequent work has focused on scientific investigations and engineering of the more advanced observational modes.  Figure~\ref{fig:photo} is a photograph of the instrument.

 PTI builds on experience gained with the Mark~III interferometer, which was a visible-wavelength interferometer installed at Mt.\ Wilson, California. The Mark~III used active fringe tracking for wide-angle astrometry and parametric imaging (\cite{M3OI}; \cite{ARAA92}).  PTI differs from the Mark~III in that it operates in the infrared, and in addition to parametric imaging supports narrow-angle interferometric astrometry implemented using phase referencing, described below.

 \section{Narrow-angle interferometric astrometry}
 \label{sec:narrowangle}

The development of technology for detecting extrasolar planets is one of the motivations for building PTI.   Most of the recent discoveries of extrasolar planets have been through measurement of radial velocities (\cite{Mayor}).  This technique senses the doppler shift of the light from a star caused by the longitudinal component of an unseen companion's motion.  A complementary technique to radial velocities is astrometry, which measures the wobble in the position of a star caused by the transverse component of a companion's motion.  While high-accuracy global astrometry remains the purview of space missions, the detection of extrasolar planets through astrometry is primarily a narrow-angle problem, as the measurement can be made with respect to angularly-nearby reference stars.  In this case the atmospheric limits to astrometric accuracy achievable with a long-baseline interferometer are several orders of magnitude below the wide-angle atmospheric limit (\cite{AA92}; \cite{AA94}). Figure~\ref{fig:atmos} shows the increase in astrometric accuracy achievable in a differential measurement with long baselines and narrow fields. For interferometer baselines of $\sim$100~m, and star separations of 15--20~arcsec, accuracies of tens of microarcseconds in an hour measurement are possible.

 To exploit this accuracy, both the parent star of the putative planet (henceforth primary star) as well as one or more astrometric reference stars (henceforth secondary stars) must be detectable by the interferometer. For the nearby sample, for which the astrometric technique is most sensitive, the primary star will be bright and provide an adequate signal-to-noise ratio within an $r_0^2\tau_0$ coherence volume to track.  However, angularly-nearby stars will generally be too faint to fringe track.  But for secondary stars chosen within the isoplanatic patch of the primary star, the primary star can be used as a phase reference (\cite{ARAA92}; \cite{Phase-ref}) to synthesize a long coherent integration time, greatly increasing sensitivity.  Phase referencing works best in the near-infrared, which provides a large isoplanatic patch and the ability to exploit large apertures.  Combined with adaptive optics, which corrects wavefront aberrations across the individual pupils, the fringe stabilization provided by phase referencing allows for sensitivities similar to those of a space-based system (\cite{ARAA92}).  Sky coverage---the fraction of the sky over which a suitable phase reference is available---is a key metric.  PTI, as a testbed with moderate apertures, has limited sky coverage.  However, good sky coverage can be achieved with apertures in the 1.5--2.0-m range, as proposed for the Keck Interferometer outriggers (\cite{Keck}).

Combining phase referencing with simultaneous measurements of two stars is the basis of dual-star astrometry.  Figure~\ref{fig:dualstar} shows the basic implementation.  Light from separate stars in the field-of-view of each aperture is directed into separate interferometers sharing a common baseline.  Internal path delays for each interferometer are measured with laser metrology to common fiducials at each aperture.  These common fiducials ``tie together'' the two interferometers, allowing, for example, fringe-tracking errors on the primary star to be fed forward to the secondary star to implement phase referencing.  The secondary interferometer, cophased by the primary interferometer, makes the actual astrometric measurement by switching between the secondary star and the primary star; the change in delay between the two stars is the astrometric observable.

Because of the small field-of-view for the narrow-angle measurements, requirements on baseline knowledge are modest.  The error $\delta$ in a differential astrometric measurement over an angular separation $\theta$ caused by a fractional baseline error of $\delta B/B$ is attenuated by the separation as $\delta = \theta (\delta B / B) $. For example, a 100~\um\ error in a 100~m baseline for astrometry over a 20~arcsec separation introduces errors of only $\sim$20~\uas.  Thus, separate metrology of the baseline is generally unnecessary, and baseline solutions can be obtained from wide-angle astrometry using one of the interferometers (\cite{M3astrometry}).

 \section{Instrument overview}
 \label{sec:overview}

PTI is located northeast of the 5-m Hale Telescope at Palomar Observatory, Palomar Mountain, California; Figure~\ref{fig:photo} is a photograph of the interferometer taken from the catwalk of the 5-m telescope.  The central building houses the optical delay lines, beam-combining laboratory, and the computer/control room.  The three small shelters located north, south, and west of the central building contain the siderostats, telescopes, and dual-star feeds.  Light is directed from the siderostat shelters through evacuated pipes to the central building.

PTI is operated as a two-element interferometer, combining light from two of the three siderostats at one time.  Initial observations combined light from the north and south piers for a 110-m baseline which is oriented 20\arcdeg\ east of north.  The inclusion of the west pier allows additional baselines. Table~\ref{tab:PTIstats} summarizes some key instrument parameters.
 Figure~\ref{fig:beamtrain} is a schematic of the beam train of PTI. Light at each aperture is collected by a 50-cm siderostat mirror, which directs the light to a fixed 40-cm telescope; a wide-angle camera adjacent to the telescope provides coarse star acquisition to get the stars within the field-of-view of the angle tracker in the beam-combining lab. The 40-cm telescope forms an image of the field, which is split into primary and secondary beams by the dual-star feed; fast-steering mirrors in the dual-star feed correct for wavefront tilt.  The two beams from each aperture are directed through beam pipes to the central building where they are directed into a common long delay line, providing identical delays for the two beams. Local metrology of the delay line is used for servo control.  Beam compressors reduce the beam diameters to a more convenient size before sending the primary star through a short delay line.  This delay line introduces a differential delay between primary and secondary to enable phase referencing within the isoplanatic patch.
       PTI has separate beam combiners for the primary and secondary stars. In each the visible light (0.7--1.0~\um) is split off with a dichroic beamsplitter and directed to an angle-tracker which controls the fast-steering mirrors in the dual-star feeds.  The infrared light (1.6--2.4~\um) is combined in a Michelson beam combiner feeding one 128$\times$128-pixel quadrant of a NICMOS-3 array.  In dual-star mode, ``constant-term'' metrology (Sec.~\ref{sec:ct} monitors the entire optical path from beam combiner to dual-star feed.  A distributed real-time control system orchestrates operation of the system.

 \paragraph{Design considerations}

Operation at 2.2~\um\ is optimal for dual-star astrometry (\cite{AA92}), exploiting both the large coherence volume and increased isoplanatic patch available in the infrared.  While the choice of a 40-cm aperture for PTI was established primarily on the basis of cost, it is also well matched to average atmospheric conditions.  For nominal 1~arcsec seeing, the coherence diameter $r_0$ at 2.2~\um\ is 53~cm; thus a $d$ = 40-cm aperture is well corrected with only modest tip/tilt correction. 

The PTI design incorporates most of the usual system requirements.  The layout is highly symmetric in order to minimize visibility losses due to differential image rotation and differential phase effects.  To maintain high throughput, most optics use protected-silver coatings.  Transmissive optics are generally antireflection-coated IR-transmissive fused silica, which work well out to 2.5~\um.   Dust covers are used in most places to protect the optics, and a CO$_2$-snow gun is used to clean the siderostat mirrors.

Atmospheric dispersion is relatively small over the K~band (App.~\ref{sec:disp}), and thus air delay lines can be used, allowing for considerable simplicity of implementation.  However, because of the long propagation paths in the delay lines and in beam transport, care must be taken to minimize turbulence, as discussed below.

The system was designed to be highly modular.  This allowed subsystem assembly and testing prior to installation at the site.  For example, the pointing system was debugged on the roof of our lab at JPL, while closed-loop operation of the fringe tracker with the delay line was accomplished in the basement of the same building.  This modularity extended to the computer system, where subsystems were allocated their own real-time computer board, even if computational needs did not mandate it, in order to simplify the development process.

 \section{Building and site}
 \label{sec:building}

The central building is a T-shaped modular insulated building with a length of 30.5~m. The top part of the building, 6.1~m $\times$ 18.3~m, houses the beam-combining lab and computer room; the latter is partitioned off from the rest of the building and located adjacent to the driveway in Fig.~\ref{fig:photo}. Figure~\ref{fig:building} gives more detail.

 The beam-combining lab includes three optical tables.  These optical tables are grouted onto 0.45-m--diameter concrete posts on top of a central isolation pad. The center 1.2~m $\times$ 3.7~m table supports relay optics, beam compressors, metrology components, and the short delay lines.  The other two 1.2~m $\times$ 2.4~m tables support 1.2~m $\times$ 2.4~m $\times$ 0.1~m breadboards on which are assembled the primary and secondary beam combiners; the use of breadboards simplified integration at Palomar, as most assembly could take place at JPL. 

The long part of the building, 4.9~m $\times$ 24.4~m, encloses the long delay lines.  Two 22~m $\times$ 0.6-m--deep I-beams provide the primary delay-line support.  They attach via embedded J-bolts to the central isolation pad at the front and at 3 other isolated pads along their length.  Box-beam crosspieces, 0.10~m $\times$ 0.15~m $\times$ 1.7~m, are located at 0.6~m intervals along the I-beam.  Each delay line runs on a pair of 38.1-mm--diameter round steel rails with extruded aluminum supports which mount to the crosspieces; a center square rail is used for the friction drive. The support system allows for three delay lines, although only two are installed.

A ``building-in-a-building'', an approach first used on the SUSI interferometer (\cite{SUSI}), is implemented using 1.2~m $\times$ 2.4~m $\times$ 0.05~m rigid insulating panels surrounding the optical tables and long delay lines as shown in Fig.~\ref{fig:building}. The building-in-a-building allows the heating and air-conditioning units, which are mounted on isolated pads, to run continually, maintaining a constant interior temperature without affecting internal seeing. Additional thermal-control measures include the use of  white paint on the building driveway and adjacent road; white-gravel is used in the vicinity of the siderostat shelters.

The siderostat shelters house the siderostats, telescopes, and dual-star feeds.  They are insulated buildings, 5.2~m $\times$ 2.7~m, elevated 1.5~m above the local grade.  With the exception of one fixed wall, the roof and upper wall roll back on supports, as seen in the foreground shelter in Fig.~\ref{fig:photo}, leaving a 0.75-m--high lower wall.  In operation, a windscreen made of rigid insulating panels surrounds the interior optics, and  a separate cover encloses the dual-star feed.  The shelter includes an electronics rack for the remote electronics, which is vented to the outside.

The pipes that bring light from the siderostat shelters to the central building are evacuated tubes capped with IR-transmissive fused silica windows. They terminate approximately 1-m from the siderostat shelters.  During observations, when the shelter roofs are rolled back, short lengths of PVC pipe enclose the light paths from the end of the evacuated pipes to the edge of the dual-star feed cover to minimize local seeing. Similarly, at the termination of the evacuated pipes at the walls of the central building, short lengths of PVC pipe enclose the light path into the building-in-a-building.

We note that the light pipes were originally simple non-evacuated insulated aluminum tubes.  Sufficiently insulated, these tubes worked acceptably for propagating the 2-\um\ stellar beams.  However, as part of a system upgrade in 1997 (\cite{kent}), the tubes were replaced with evacuated pipes, which improved the propagation of the 633-nm constant-term metrology beams, which make a double-pass through the tubes.  The driving disturbance in the air pipes was the temperature difference between the tops of the pipes, radiatively cooling to the sky, and the bottoms, which would be coupled to the warmer ground.  While the insulation minimized the effect, some convection was seen at dusk when setting up the instrument; we suspect that the air pipes would be entirely acceptable if buried and away from diurnal disturbances.

 \section{Siderostat system}
 \label{sec:siderostat}

The siderostat mount is an alt-az gimbal moving a 50-cm zerodur flat. The mirror is supported on a 9-point wiffletree suspension.  Adjustable hard-points, opposed by spring plungers, locate the mirror surface with the altitude axis.  The altitude bearing mounts can be shimmed to align the altitude and azimuth axes.  The mount provides access to the back surface of the siderostat mirror through a hole in the mirror cell.  A contacting depth gauge, encoded by a laser interferometer, was used to measure from a fixed point off the mount to the front or back of the siderostat mirror with the mount rotated to (az, el) coordinates of (0$\arcdeg$, 0$\arcdeg$), (180$\arcdeg$, 0$\arcdeg$), (0$\arcdeg$, 180$\arcdeg$), and (180$\arcdeg$, 180$\arcdeg$).  Combined with the center thickness of the siderostat mirror, these measurements allow determination of the offset between the two axes and the mirror surface offset.  (As the measurements are always made at near-normal incidence to the front or back surface of the mirror, tolerances on the precise mount angles at which measurements are made are relatively loose.)  The mount is adjusted to keep deviations from an ideal siderostat to $<$100~\um.  Mirror edge location is via delrin tabs, and the unbalanced altitude load is compensated with a counterweight.

The system is driven with a dc servo motor.  It provides 0.027~arcsec motor steps and a maximum slew rate of $\sim$1\arcdeg/s.  Direct axis encoding is  provided by 6.48-arcsec optical encoders; the zero points of each axis are defined by magnetic proximity switches on the main wheel repeatable to less than one encoder step. Computer readable ``soft'' limit switches indicate extremes of travel while ``hard'' limits protect against software failures by disabling the siderostat drive.

 \paragraph{Control aspects}

The real-time control system for the siderostats provides track and slew control based on sidereal targets from the instrument sequencer, as well as variety of other target types. Slew and track control use rate feedback to the axis indexers: large-angle slews use the encoders for closed-loop control, while tracking uses the indexer position; transition between modes is automatic based on the size of the tracking error.  The controller accepts offsets from the acquisition cameras and angle trackers for star acquisition, and from the angle tracker for fast-steering-mirror desaturation. The translation from sidereal targets to mount position is via an 8-parameter mount model which incorporates the orientation of the axes, the zero points of the axes rotations, the orientation of the siderostat mirror in its cell, and the direction of the feed beam from the telescope. These parameters are determined via an off-line least-squares routine using pointing residuals to known targets; when properly calibrated, open-loop pointing is good to $\sim$15~arcsec rms---well within the field-of-view of the acquisition system.  The only nightly calibration required of the siderostat system is a homing process to establish the encoder zero points.  Manual paddle controls are provided but are not required in normal operation.

For observations in 1997 an accelerometer feedforward system was implemented to compensate for a small amount of piston attributable to the siderostats which was detectable in the fringe phase.  The system uses an accelerometer\footnote{Sunstrand QA3000}
 mounted to the back of each siderostat mirror which senses acceleration along the mirror normal.  Analog electronics filter and doubly integrate the accelerometer output to provide a measure of the mirror piston above 1.5~Hz.  This signal is digitized at 2~kHz, scaled by $2\cos\theta$, where $\theta$ is the angle between the siderostat normal and the telescope feedbeam, and is fed forward to the delay line for compensation. 

 \section{Acquisition system} 
 \label{sec:acq}

The siderostat mirror directs light to a 40-cm Cassegrain telescope. However, preceding the telescope, an elliptical mirror located in the shadow of the telescope secondary directs a fraction of the stellar beam to a wide-angle star acquisition system, as shown in Fig.~\ref{fig:acq}. Given the long baselines and high magnification of the beam-transport system, accurate pointing is required in order for starlight to enter the field-of-view of the angle tracker inside the beam-combining lab.

 The acquisition system uses a commercial 12-bit thermoelectrically-cooled CCD camera\footnote{Photometrics Star-1: format: 384$\times$576; pixel size: 23~\um; read noise: 20~e$^-$; dark current: 15~e$^-$/s at -45~C}
 which interfaces to the control system using GPIB. The camera is preceded by a 90-mm diameter, 1300-mm focal-length telescope followed by a 2:1 teleconverter.  These provide an angular scale of 1.8~arcsec/pixel over a field-of-view of 12x18~arcmin; smaller fields are selectable via subframe readout.

The acquisition system incorporates a 70-mm corner cube for calibration. The corner cube is located at the edge of  the elliptical mirror feeding the acquisition system so that a portion of a boresight laser injected in the beam-combining lab and fed backwards through the system is intercepted and directed into the acquisition camera.

 \paragraph{Control aspects}

The acquisition system implements a basic centroid operation to locate the star image in the field, using a 10$\times$10 box about the brightest pixel.  The exposure time is optimized if the intensity if out of defined bounds.  Subframe readout is used after the first exposure to minimize image-transfer time.   Feedback of centroids to the siderostats is used to automate the acquisition process. The acquisition system also provides information on apparent star magnitude to the angle tracker.

The acquisition system provides several calibration modes. A {\it primary boresight}, done at the beginning of a night, records the image centroid of the primary boresight laser retroreflected off the acquisition corner cube; this location establishes the initial zero point for star centroiding.  This zero point is updated between scans, once the angle tracker has acquired the star, using a {\it star boresight}, which records the current stellar position. The repeated star boresights remove drifts in the acquisition system as well as any systematic offsets due to aberrations in the primary laser image.  For dual-star operations a {\it secondary boresight} measures the image centroid of the secondary boresight laser.  This is used to servo the secondary star selector mirror in the dual-star feed to coalign the secondary and primary boresights.

 \section{Collector telescope and dual-star feed}
 \label{sec:dsf}

The siderostat mirror feeds a 40-cm Cassegrain telescope. The telescope is mounted to a 1.2~m $\times$ 2.4~m optical table, inclined 15\arcdeg\ with respect to horizontal; the telescopes point 15\arcdeg\ north of west at the siderostats (\cite{PTI94}).  The north and south telescopes employ a thermally-compensated truss (\cite{Krim}) with invar truss elements and aluminum end pieces, while the west telescope uses a rail-mounted secondary which is athermalized with an invar metering rod.  The telescopes use an f/3 primary and provide an f/10 output beam at a back focal distance of 400~mm.  The secondary diameter is 124~mm, with a secondary obscuration of 140~mm.  The wavefront quality of the telescopes is 30--40~nm rms.  The 15-degree inclination of the telescopes improves the vignet-free field-of-regard of the siderostat.  For sources at a declination of $\delta$ = 30\arcdeg, the system is unvignetted for hour angles HA $<$ 0; areal vignetting remains $<$10\% for HA $<$ 1.5~h and $<$20\% for HA $<$ 2.5~h. Figure~\ref{fig:sky} provides additional information on the field-of-regard of the system.

The telescope feeds the dual-star feed (DSF), shown schematically in Fig.~\ref{fig:dsf}. The DSF separates the light from the primary and secondary stars into separate beams and directs them to the beam-combining lab. The DSF also contains the fast-steering mirrors (FSMs) used to correct wavefront tilt. The DSF is constructed on a 0.9~m $\times$ 1.2~m optical breadboard mounted horizontally behind the telescope.  

With reference to Fig.~\ref{fig:dsf}, mirror M1 folds the inclined output beam of the telescope into the horizontal plane where it is focused onto the field separator M2 at a scale of 19.4~\um/arcsec.  The primary beam transmits through the field separator and is reflected by mirror M3 through a 25-mm hole in mirror M5 to parabola M4.  The parabola has a focal length of 750~mm and recollimates the starlight into a 75-mm beam (compression of 5.33 with respect to the telescope primary).  The collimated beam reflects off annular mirror M5 to fold mirror M6, which directs the primary beam toward the beam-combining lab.

The large 75-mm collimated beam minimizes diffraction in the beam transport and also reduces delay-dependent phase and amplitude errors (\cite{Diffraction}).  The Rayleigh length $d^2/\lambda$ for propagation at 2.2~\um\ is 2600~m; with $\sim$100~m internal propagation distances, PTI operates at a small fraction of this value.

While the primary beam transmits through the field separator M2, the secondary beam is reflected, and is recollimated and directed similarly to the primary beam with mirrors M4\p, M5\p, and M6\p. Mirror M6\p\ is referred to as the secondary star selector mirror (SSSM), and is located conjugate to the plane in the 40-cm beam between the siderostat and telescope which includes the telescope secondary.  This plane defines the system input pupil, and also includes (physically or virtually) the constant-term fiducials, discussed in Sec.~\ref{sec:ct}. Tilt of the SSSM is used to select a secondary star from the telescope image on M2. An optional Lyot stop can be located adjacent to the SSSM.

The SSSM is a flexure mount driven by two microstepped actuators; including the 5.33 demagnification and a factor of 2 for reflection, its angular resolution on the sky is 0.015~arcsec.  The SSSM is controlled by the siderostat control system, tracking the sidereal motion of the secondary star with respect to the primary star and accepting offsets from the acquisition and angle-tracker systems.

Mirrors M4 and M4\p\ are also flexure mounts and are driven with 15-\um\ PZT actuators for control of high-frequency tip/tilt sensed in the beam-combining lab.  They have a range on the sky of $\pm$10~arcsec. When tracking, only a small portion of this range is used.

Several options are available for the field separator M2. For single-star observations, the field separator is just removed.  For dual-star observations, and especially for engineering tests, an achromatic beamsplitter is convenient.  For application in the converging beam we use a dielectric deposited on a 300-\um--thick substrate.  For large magnitude differences between primary and secondary, the optimal field separator is a pinhole.  However, to maintain metrology continuity, the pinhole must be deposited on a substrate incorporating a metrology beamsplitter coating. 

 \section{Central optics and delay lines}
 \label{sec:central}

Figure~\ref{fig:detail} is a schematic of the central area of the beam-combining lab. The 75-mm collimated beams from the primary and secondary stars enter the central building as shown in Fig.~\ref{fig:building}.  The light is directed by feed mirrors on the center table into the long delay lines, which occupy the long part of the central building.

 \subsection{Long delay lines}
 \label{sec:bigdl}

The long delay lines use a four-stage design nearly identical to one previously developed at JPL (\cite{dl1}; \cite{dl2}). The optical assembly is an athermalized cat's eye implemented with a parabola and a flat. The parabola is 330$\times$230~mm with a 1070-mm focal length. The flat at the system focus is mounted on a PZT for high-bandwidth servo control; a compensating PZT provides momentum balance.  The optical assembly is suspended with flexure arms above an optics cart that rolls on steel wheels on the delay-line track.  The optical assembly is driven with respect to the optics cart with a voice-coil motor for intermediate-level control.  A separate motor cart contains a microstepper motor which provides coarse motion using a friction drive to the center rail; a second voice-coil motor drives the optics cart with respect to the motor cart.  Control and drive signals are provided via a ribbon cable which runs in a cable tray from the front of the delay-line track, under the delay-line cart, around an idler pulley, and then to the motor cart.  The idler pulley is on a third cart, which is tensioned via a cable from a torque motor at the back of the delay-line track.  The delay line maintains delay jitter of 10--20~nm rms at speeds of tens of mm/sec. This is well in excess of PTI requirements of jitter $<$ 50~nm for a 1\% visibility reduction at 2.2~\um\ at a maximum sidereal rate of 5~mm/sec.

PTI uses two long delay lines.  The south (``active'') delay line is of the type just described.  The north (``passive'') delay line includes just the motor stage and is moved only between observations.  Thus quiet motion is not required and it can use a more rudimentary (and less expensive) optics cart that rides on ball bushings and has rigid connections to both the optical assembly and the motor cart.  Residual motion of the passive delay line during an observation is sensed by laser metrology and servoed out by the active delay line. The delay-line tracks are 22-m long; allowing room for the various carts and some margin for deceleration from slew, the physical travel of each delay line is 19.15~m for a total optical delay using both delay lines of $\pm$38.3~m.  The delay-line travel was chosen subject to cost constraints, and sets the limits of sky coverage on the N-S baseline, shown graphically in Fig.~\ref{fig:sky}.

The delay lines incorporate a PZT for modulation which is mechanically in series with the servo PZT.  This PZT is driven with a fringe-scanning waveform as required for fringe detection. The use of a separate PZT for modulation allows for simple distortion correction; a waveform-generator card provides a predistorted drive waveform to correct for PZT nonlinearities.  If the modulation were implemented as an additional signal on the servo PZT, its nonlinearity would be a function of the operating point of the servo.
 
 The long delay lines delay both the primary and secondary stars.  With respect to the parabolic mirror, 75-mm primary and secondary beams enter at the upper left and upper right, and exit diagonally opposite the optical axis at the lower right and lower left.  These delayed beams exit the delay lines and pass beneath the feed mirrors toward the beam compressors.

 \subsection{Beam compressors}

The beam compressors reduce the starlight beam diameter by a factor of 3.75 from a transport diameter of 75~mm to 20~mm to reduce the size of the remaining back-end optics.  They use an on-axis Gregorian design with confocal parabolas. The primary is 100~mm in diameter with a 500-mm focal length.  The compressor despace is maintained with invar rods for athermalization. The projected obscuration of the compressor is less than that of the main telescope secondary.

 \subsection{Short delay lines} 
 \label{sec:shortdl}

After the beam compressors the primary beams pass through the short delay lines.  These introduce a differential delay between the primary and secondary star.  Thus, if $L$ and $l$ are the delays introduced by the long and short delay lines, the delays seen by the primary and secondary stars are $L + l$ and $L$, respectively.  With this delay-line configuration, pathlength modulation can be implemented on the long delay line, affecting both the primary and secondary stars, or on the short delay line, affecting only the primary star.

The requirements on the small delay line are modest: for a $\pm$30~arcsec range on the sky with a 110-m baseline, the required delay is $\pm$17~mm with a maximum sidereal rate of 1~\um/s. The south small delay line is active, with a PZT and motor, while the north one is completely passive and included for optical symmetry. The small active delay line implements a simple two-stage servo.  It uses a PZT at the focus of an f/5 100-mm parabola and a microstepper-driven translation stage which moves the entire assembly.  

 \subsection{Beam combiner tables} 

The 20-mm primary beams delayed by the short delay line are directed to the primary beam-combining table; the light from the secondary star is directed after its compressors directly to the secondary beam combiner. The beam-combiner tables include the optics and detectors for the angle tracker and fringe tracker.  Figure~\ref{fig:primary} is a schematic of the primary beam combiner.  The secondary table is similar, but uses a different fringe-detector configuration. The angle tracker uses light reflected from dichroic B1 into telescope L1, which feeds the angle-tracker detector.  The fringe tracker demodulates the interference fringe formed at beamsplitter B3. The angle tracker and fringe tracker are discussed in detail in Secs.~\ref{sec:at} and \ref{sec:ft}, below.  Additional components on the beam combiner table include a white-light source, which is collimated and injected into the combined beam via beamsplitter B5.  The white-light source is used for boresighting and calibration of both the fringe tracker and angle tracker.  A HeNe laser can be injected via beamsplitter B6 and is used for boresighting the acquisition system, as discussed in Sec.~\ref{sec:acq}, as well as for manual boresighting and focusing.  A microscope at the back of the system is useful for system alignment and test.  Also indicated on the figure are the locations of the constant-term metrology injection and extraction optics (see Sec.~\ref{sec:ct}).

 \section{Angle tracker}
 \label{sec:at}

The angle tracker senses and controls the wavefront tilt of the interfering starlight beams.  The system uses an angle sensor in the beam combiner that controls fast-steering mirrors in the DSF; identical systems are implemented on both the primary and secondary beam combiners.  Angle sensing is over the bandpass 0.7--1.0~\um\ using a quad-cell detector implemented with photon-counting Si APD detectors; a chopping scheme is implemented which shares the angle sensor between the two input beams on each beam combiner.  The bandpass is limited at long wavelengths by the bandgap of the Si detectors, and at short wavelengths by a long-pass filter which blocks scattered HeNe metrology light.

 \paragraph{Atmospheric issues}

Assuming a standard Kolmogorov turbulence model, the total wavefront variance across an aperture of diameter $d$ is $1.03(d/r_0)^{5/3}$ rads$^2$; with perfect tip/tilt correction, the variance is reduced to $0.134(d/r_0)^{5/3}$ (\cite{Noll}). If the angle tracker removes 90\% of the tip/tilt variance, the residual wavefront error for one aperture is $\sigma$ = 0.37~rads rms.  This results in a mean value for the squared fringe visibility of $\exp{(-2\sigma^2)}$ = 0.76 (see \cite{TangoTwiss}).   To achieve this performance requires only modest angle-tracker bandwidths. For a wavefront tilt power spectrum $H(f)$, and wind speed $W$, the product $fH(f)$ peaks at 0.075--0.25$W/d$ depending on whether the wind is transverse or longitudinal (\cite{Roddier}); most of the tilt energy lies below these frequencies. Assuming $W$ = 10~m/s and $d$ = 40~cm, the peak is at 2--6~Hz.

 \paragraph{Description}

The angle tracker components are mounted on the beam combiner tables as shown in Fig.~\ref{fig:primary}.  The 20-mm collimated beams from each aperture enter from the left.  Dichroic B1 reflects the bandpass from 0.7--1.0~\um\ and transmits at 633~nm and 1.5--2.4~\um.  The reflected light from B1 is directed to telescope L1 (diameter = 90~mm; focal length (fl) = 1300~mm) which forms non-overlapping images of each beam on field lens L2 (fl = 38~mm). Mirrors M1 and M1\p\ are adjusted such that the beams cross at plane R1  prior to the field lens L2; this plane is imaged by L2 via fold mirror M2 onto mirror M3. Lens L3 (fl = 250~mm) relays the image from field lens L2 onto the lenslet array L4 which defines the 4 quadrants of the detector.  L4 is portion of a commercial\footnote{Adaptive Optics Associates}
 replicated microlenslet array; the lenslet pitch is 400~\um\ with individual focal lengths of 3.2~mm.  The lenslet is mounted to the front of a custom assembly containing 4 100-\um\ optical fibers at the focus of 4 of the lenslets; the fibers, with opaque jackets, feed the 4 detectors. Filter F1 is a long-pass filter which blocks scattered laser light at 633~nm.

The detectors are commercial\footnote{EG\&G SPCM-200-PQ}
 passive-quenched avalanche photodiode (APD) photon counters selected for low dark count ($<$5 counts/s). They provide a raw quantum efficiency of $\sim$27\% across 0.7--1.0~\um\ and are usable to count rates of $\sim$1~MHz. Both beams share the detectors. This is accomplished by mounting mirror M3 on a PZT tip/tilt actuator. M3 is driven by a 100-Hz square-wave signal, alternately placing images of each beam onto the lenslet array L4.   The image scale at L4, referred to the sky, is 130~\um/arcsec, and thus the lenslet array defines a quad-cell with a size of 6$\times$6~arcsec.  The separation of the chopped images is 8~arcsec.

 \paragraph{Control aspects}

The fundamental frame time of the system is 10~ms. Each frame includes a 4.3-ms integration on each beam; the dead time allows for settling of the chopper. Computer interface is through gated pulse counters.  The fundamental track controller implements a conventional quad-cell algorithm and provides integral feedback to the FSMs in the DSF; control bandwidths to $\sim$10~Hz can be implemented. Lower-bandwidth feedback is provided to the siderostat (or SSSM) for FSM desaturation. For faint sources, coadding of 10-ms frames can be selected.  A spiral search algorithm, through direct handoff to the siderostat (or SSSM for the secondary angle tracker) is used for star acquisition.  The search algorithm spirals out and then back to the origin using increasing search radii; the return to origin makes the algorithm robust against clouds and other beam blockages.  The measured source intensity from the acquisition system is used to establish a detection threshold.

The secondary angle tracker can also accept direct high-bandwidth tip/tilt feedforward from the primary angle tracker in order to implement an isoplanatic tilt correction from the primary star to the secondary one.  This can be implemented in conjunction with low-bandwidth tracking on the secondary star itself to take out slow drifts.

Calibration of the angle tracker uses two corner cubes to generate reference images.  The white-light source injected at beamsplitter B5 propagates backwards through the system.  It generates two beams which reflect off dichroic beamsplitter B1, retroreflect from corner cubes C2 and C2\p, and then transmit through B1 to produce images on the quad cells. The angle tracker adjusts the end-points of the chop waveform driving mirror M3 in order to center each image at the quad-cell center.  This {\it chop calibrate} procedure is one of our initial nightly calibrations.

The angle tracker also includes support for other less frequent calibrations, including measurement of detector quantum efficiency and calculation of the coordinate transformation matrices between detector and FSM, detector and siderostat (or SSSM), and FSM and siderostat.

 \section{Fringe tracker}
 \label{sec:ft}

The fringe trackers on PTI measure the fringe parameters of the interfering light.  The fringe phase is used to control the optical delay lines to track the atmospheric fringe motion.  The fringe phase from the primary star can also be used to cophase the secondary star. More precisely, three modes are supported:

 \begin{description}

 \item[single star]  In this case, the secondary channel is disabled.  The primary fringe tracker works at high speed to follow the atmospheric motion.  The short delay line is disabled, and control handoffs are given to the long delay line.  While the nominal sample time is 10~ms, coadding adjacent samples can be used, under conditions of good seeing, to increase the signal-to-noise ratio.

 \item[bright secondary]  This case adds fast tracking of the secondary fringe motion.  Delay-line handoffs are trivially orthogonalized to allow independent tracking.  As discussed in Sec.~\ref{sec:shortdl}, the long and short delay lines both contribute to the primary delay, while only the long delay line contributes to the secondary delay.  Thus, if $\widehat{d_p}$ and $\widehat{d_s}$ are filtered delay updates for the primary and secondary, delay-line handoffs to the long and short delay lines are generated as $\delta L = \widehat{d_s}$ and $\delta l = \widehat{d_p} - \widehat{d_s}$.

 \item[faint secondary]  In this phase-referenced mode, high-bandwidth control of the secondary fringe position comes from the primary fringe tracker, while low bandwidth control comes from the secondary star itself.  Delay-line handoffs are orthogonalized as $\delta L= \widehat{d_p} + \widehat{d_s} $ and $\delta l = -\widehat{d_s}$.

 \end{description}

 \subsection{Atmospheric issues}

PTI uses coherent tracking of the white-light interference fringe, with lower bandwidth fringe centering obtained from the group delay. Continuity requirements applied to the modulo-$2\pi$ fringe-phase measurements set the sample spacing.  For 1-arcsec seeing, corresponding to a coherence diameter $r_0$ = 53~cm at 2.2~\um, and a constant wind speed $W$ = 10~m/s, the fringe structure function is equal to one radian rms at coherence time $\tau_{0,2}$ = 11~ms (see App.~\ref{sec:phase}).  For a simple phase unwrapper, the sample spacing must be less than this value to minimize the probability of unwrapping errors, and we adopt a value of $t=$10~ms.  With a more sophisticated phase unwrapper, as implemented on PTI, the effective coherence time increases to $\tau_{0,2}^{\prime}$ = 14~ms (see App.~\ref{sec:kf}). 

Fringe motion during the coherent integration time reduces fringe visibility squared as (\cite{TangoTwiss}): $\exp{(-T/T_{0,2})^{5/3}}$.  The quantity $T_{0,2}$ is the coherence time associated with fluctuations about the interval mean (see App.~\ref{sec:phase}), given by $T_{0,2} = 0.815r_0/W$ = 43~ms.  For $T$ = 6.75~ms (discussed below), the reduction in $V^2$ due to temporal effects is 5\%.   Finally, the two-aperture Greenwood frequency can be computed as $f_{G,2} = (1.546r_0/W)^{-1}$ = 12~Hz. For a servo bandwidth $f_c$ = 10~Hz (1/10 of the 100-Hz sample rate), the residual fringe-tracker error is $(f_{G,2}/f_c)^{5/6}$ = 1.2~rads rms.

 \subsection{Optical configuration}

The optical configuration of the primary beam combiner is shown in Fig.~\ref{fig:primary}. Dichroic beamsplitter B1 passes the infrared light from the two 20-mm input beams, and a flat fringe is formed at beamsplitter B3; compensator B2 accounts for the thickness of the beamsplitter substrate.  The combined light from one side of the beamsplitter (``white-light output'') is directed via lens L5 (200~mm fl) and mirrors M5 and M6 to an external focus adjacent to mirror M10.  The light from the other side of the beamsplitter (``spectrometer output'') is directed via dichroic beamsplitter B4 through a spatial filter assembly and external spectrometer to form a line focus on mirror M10 adjacent to the white-light spot.

Dichroic beamsplitter B4 is 90\% reflective from 1.5--2.4~\um; the residual transmission allows injection of the white-light and laser sources in the combined beam for calibration. The spatial-filter assembly is implemented using off-axis portions of parabolas M7 and M8 which focus the light into a single-mode fluoride-glass fiber\footnote{Le Verre Fluore: $\lambda_c$ = 1.45~\um, NA = 0.176.}
 and recollimate the exiting light.\footnote{The spatial filter was used for all observations in 1997, and several observations in 1996.}
 As the spatial filter is implemented in collimated space with mirrors M7 and M8 on kinematic bases, spatial filtering can be removed if desired with only minor realignment of the system.  To maximize tracking sensitivity, no explicit spatial filtering is used on the white-light channel, although some filtering results from the finite pixel size.

Prism P1 and lens L6 (200~mm fl) implement a low-resolution spectrometer.  Prism P1 can be chosen as desired, but is typically a simple 8--12\arcdeg\ (physical) fused silica wedge.\footnote{The original spectrometer configuration used a single 16-deg fused silica prism; the configuration for 1997 reduced the dispersion with a -7.68-deg BK7 prism between M9 and L6; the text describes the current configuration.}
 Lens L6 (like L5, L7, and L8) is a custom air-spaced achromat (fused silica $+$ calcium fluoride), afocal from 1.5--2.4~\um.

The external focus at M10 is recollimated with lens L7 (100~mm fl) and relayed into a nitrogen dewar.\footnote{IR Labs ND8}
 The dewar contains a filter wheel in collimated space with K (2.00--2.40~\um), K$^\prime$ (2.00--2.33~\um), and HK (1.5--2.40~\um) filters, plus a blank-off position.  Pupil stops, nominally 10-mm diameter, are attached to the filters.  A second 100-mm lens after the filter wheel relays light onto the infrared detector.

The primary beam combiner, discussed so far, works at atmospheric rates.  As it will typically be read-noise limited, the warm, slit-less spectrometer design described above is adequate.  For the secondary beam combiner, phase referencing allows longer integration times and the infrared background becomes more important.  Thus a cold spectrometer is implemented, which moves some of the functions shown in Fig.~\ref{fig:primary} into the secondary dewar. Figure~\ref{fig:secondarydewar} gives a schematic of the secondary dewar.  Light from the white-light and spectrometer outputs of the beamsplitter are separately focused onto a slit behind the dewar window by 200-mm lenses L1 and L1$^\prime$, identical to those used on the primary beam combiner.  An external in-line spatial filter assembly can be used on the spectrometer side as on the primary beam combiner.

The entrance slit, nominally 40~\um\ wide, widens to 400~\um\ for part of its length. The white-light beam enters through the wide part of the slit, is collimated by lens L3, passes through a filter wheel which includes the K, K$^\prime$, and HK filters and pupil stops, and is then relayed to the detector by lens L4.  The spectrometer beam can use either the narrow or wide part of the slit.  Because it enters off-axis, the spectrometer beam is intercepted by mirror M1, is recollimated by lens L2, and dispersed by prism P1.  Mirror M2 introduces a double pass, and is oriented to produce a line focus on the back of the slit adjacent to the white-light spot.  This spectrum is then relayed through the filter wheel to the detector via lenses L3 and L4.  For use on bright stars, the spectrometer prism P1 can be removed, M2 realigned, and an external prism inserted prior to L1$^\prime$.

 \subsection{Detector and electronics}

The detector for both beam combiners is an engineering-grade NICMOS-3 infrared array. One 128$\times$128 quadrant is used.  Pixels are 40$\times$40~\um, and the external optics are configured so that the white-light spot and the spectrum are located adjacent to each other on a single line of the detector; the spectrometer resolution typically provides 5--8 pixels across the K band.

Video amplification and clock buffering are provided with commercial electronics\footnote{IR Labs}
 on pod boxes attached to the dewar.  The video output drives a 16-bit, 500-kHz A/D converter which interfaces to the real-time system through a FIFO on a custom VME board.  This board also includes a large downloadable memory that contains the clock patterns for the array. The clock patterns for real-time use are aperiodic, and are precomputed and downloaded during nightly system configuration.

The NICMOS-3 detector provides a quantum efficiency of $\sim$65\% at 2.2~\um, and a read noise for a single 2-\us\ correlated-double-sample read of $\sim$32~e$^-$.  For use on PTI, multiple consecutive 2-\us\ reads are digitally averaged to decrease the effective read noise, typically to 11--16~e$^-$, as discussed below.

 \subsection{Fringe demodulation and array clocking}

To implement fast tracking, fringe demodulation on PTI uses a fringe-scanning algorithm, similar to that employed on the Mark~III Interferometer (\cite{M3OI}), but adopted for use with array detectors.  The modulation waveform is a 100-Hz sawtooth pattern implemented on one of the delay lines.  This unidirectional scan minimizes the number of reads required to measure the fringe parameters. The array timing is aperiodic, varying the integration time per pixel such that each pixel is scanned by exactly one wavelength. 

The timing for each 10-ms sample is identical. Each sample begins with a reset of the active and adjacent lines of the array followed by a short settling time.  During this time the stroke waveform retraces from its value at the end of the previous sample. The first read begins typically 1.5~ms into the sample and measures the reset pedestal on the white-light and spectrometer pixels.  The stroke waveform begins its linear scan at this time, and successive reads occur after each quarter wave of modulation for each pixel.  Each of these (nondestructive) reads is typically an average of 16--64 consecutive 2-\us\ subreads.  With the dewar blanked off, the system at PTI achieves effective read noises for fringe demodulation of 24, 20, 16, 13, and 11~e$^-$ at 2, 4, 8, 16, and 32 subreads.  The non-white behavior most likely corresponds to 1/f noise in the detector.

The array clocking involves some overhead associated with the array reset and settling time, although this does take place during the stroke retrace.  Additional overhead comes from the shift register implementation on the NICMOS-3 chip, which does not have a clear, so that access to a previous pixel requires clocking out, reloading the shift register, and clocking back in.\footnote{The PICNIC chip, which replaces NICMOS, has clearable shift registers and is resetable by row rather than by individual pixel.}
 Fundamentally, matching the stroke to the wavelength means that the effective integration time at midband will be less than the longest integration time allowed.  The present configuration achieves an integration time of 6.75~ms at 2.2~\um\ and 7.4~ms at 2.4~\um.

With wider wavelength coverage to accommodate both H and K bands simultaneously, varying the bin widths with pixel becomes less practical.  Appendix~\ref{sec:fixed} discusses the corrections to the fringe quadratures necessary for use with fixed-width time bins.

 \subsection{Fringe acquisition and tracking}

The fast fringe-tracker algorithm is a variant of that used on the Mark~I--III interferometers (\cite{M3OI}; \cite{M1OI}). Denote the five reads during each 10-ms sample time as $z_i, a_i, b_i, c_i,$ and $d_i$, where $i = 0$ denotes the white-light pixel and $i = 1 \ldots R$ denote the $R$ spectrometer pixels.  The integrated flux in each quarter-wave time bin is calculated as $A_i = a_i -  z_i, B_i = b_i - a_i, C_i = c_i - b_i,$ and $D_i = d_i - c_i.$  The raw fringe quadratures and total flux (in units of dn) are calculated from these values as 
 \begin{eqnarray}
 X_i & = & A_i - C_i \\
 Y_i & = & B_i - D_i \\
 N_i & = & A_i + B_i + C_i + D_i.
 \end{eqnarray}
 From these quantities we can estimate the fringe phase, visibility, and  signal-to-noise ratio, but first it is necessary to correct for biases associated with the detection and readout process.

A detailed discussion of the bias correction of these quantities is given elsewhere (\cite{PTIvis}).  While explicit calibration measurements are interspersed with science observations, two initial calibrations are used to estimate the biases for use by the real-time system.  The mean values $B^X$, $B^Y$, and $B^N$ on $X$, $Y$, and $N$ are estimated from a low-level calibration with the system observing dark sky.  Applying these bias corrections yields $\widehat{X} = X - B^X$, $\widehat{Y} = Y - B^Y$, and $\widehat{N} = N - B^N$, from which an energy measure $\widehat{\rm NUM}$ is computed as
 \begin{equation}
 \widehat{\rm NUM} = \widehat{X}^2 + \widehat{Y}^2. 
 \end{equation}The squared quantity $\widehat{\rm NUM}$ contains biases attributable to read-noise and photon noise.  The detector read-noise bias $B^{\rm rn}$ is estimated from the value of $\widehat{\rm NUM}$ measured from a low-level calibration.  The photon noise bias $B^{\rm pn}$ takes the form $B^{\rm pn} = k\widehat{N}$, where $k$ is the detector scale factor in dn/e$^-$.  The factor $k$ is estimated from the increase in the bias on $\widehat{\rm NUM}$ with increasing light level.  This high-level calibration uses the white-light source in Fig.~\ref{fig:primary}, which is retroreflected back into the dewar with a corner cube placed at the exit of the primary table.\footnote{The corner cubes C1, C1\p, while optically in the proper location, do not see much infrared light, as dichroic beamsplitter B1 is nominally transmissive at that wavelength.}
 Thus $\widehat{\rm NUM}$ is corrected for these ac biases to yield
 \begin{equation}
 \widehat{\widehat{\rm NUM}} = \widehat{\rm NUM} - B^{\rm rn} - B^{\rm pn}.
 \end{equation}

Using these bias-corrected quantities, instantaneous estimates of the fringe phase, fringe visibility, and fringe signal-to-noise ratio $S$ (photon-noise limit) are computed as
 \begin{eqnarray}
 \phi & = & \tan^{-1}\frac{\widehat{Y}}{\widehat{X}} \\
 V^2  &= & \frac{\pi^2}{2}\frac{\widehat{\widehat{\rm NUM}}}{\widehat{N}^2} \\
 S^2 & = & 2 \frac{\widehat{\widehat{\rm NUM}}}{\widehat{N}}.
 \end{eqnarray}
 With read noise, the signal-to-noise ratio (SNR) of the phase estimate is just
 \begin{equation}
 {\rm SNR}^2_{\phi} = \frac{4}{\pi^2} \frac{N^2V^2}{N+4\sigma^2_{\rm cds}},
 \end{equation}
 where $\sigma^2_{\rm cds}$ is the effective double-correlated-sample read-noise variance.

The fringe tracker supports three fundamental states: {\it search}, {\it semilock}, and {\it lock}.  Transitions between states depend on the instantaneous value of the fringe signal-to-noise ratio $S^2$, as computed above, or its average value $\overline{S^2}$, which is computed as the output of a sliding boxcar filter.  A boxcar filter is selected for its finite impulse response (rather than a decaying-memory filter).  Currently, a 15-point memory is used.

In the search state the fringe tracker commands the delay line to implement a geometric spiral about the predicted fringe position.  Typically, the search step is 4.4~\um\ every other sample, with an initial search out to +50~\um, back through zero to -100~\um, etc. Transition from search to semilock occurs for $S^2 > T^2_1$.

In the semilock state the fringe tracker attempts to track the white-light fringe.  If $\overline{S^2} > T^2_2$ after a time-out period, typically 100~ms, the fringe tracker enters the lock state; otherwise, it reverts back to the search state and continues the spiral pattern.

In the lock state the fringe tracker tracks the white-light fringe. Loss of lock, and reversion to the search state and a restart of the search process about the last lock position, occurs if $\overline{S^2} < T^2_3$.  The transition thresholds $T_1$, $T_2$, and $T_3$ are typically 6.0, 4.0, and 3.3; as implemented, they incorporate a weak dependence on photon flux to accommodate uncompensated biases at high light levels.

The fringe-tracker servo is straightforward.  The white-light phase at sample $n$, $\phi(n)$, is unwrapped to produce $\Phi(n)$, which is filtered and used to update the delay-line position.  While a fixed-bandwidth tracker is employed, a Kalman-filter phase unwrapper is used to improve performance.  Let $\chi(n)$ be the atmospheric phase process; a state-space model and estimator for $\chi$ is presented in App.~\ref{sec:kf}.  The atmospheric phase is the sum of the unwrapped white-light phase plus the offset from the predicted sidereal position of the delay line, $L$, viz. $\chi(n) = \Phi(n) + kL(n)$. The white-light phase $\phi(n)$ is unwrapped about a prediction for the unwrapped phase at time $n$ as
 \begin{equation}
 \Phi(n) = {\rm mod}_{-\pi,\pi}[\phi(n) - \widehat{\Phi}(n|n-1) ] + \widehat{\Phi}(n|n-1),
 \end{equation}
 where $\widehat{\Phi}(n|n-1)$ is computed with a Kalman-filter predictor of $\chi(n)$ as
 \begin{equation}
 \widehat{\Phi}(n|n-1) = \widehat{\chi}(n|n-1) - kL(n).
 \end{equation}
 The update for the Kalman filter uses the observation $y(n)=\Phi(n) + kL(n)$.

A simple integral fringe-tracker controller is implemented using the error signal $\Phi(n)$.  Essentially, the delay-line track handoff is updated with a new rate at sample boundaries.  The fringe-tracker gain is set to achieve a closed-loop bandwidth of 10~Hz.

 \subsection{Fringe centering}

The K-band fringe envelope contains several fringes that have adequate SNR for tracking.  For accurate phase referencing and astrometry, as well as for the best visibility measurements, it is necessary to choose the fringe closest to the center of the white-light fringe envelope, i.e., to work at zero group delay.  While the group delay is readily estimated using the spectrometer channels, because of their lower SNR the group delay is not available at the full sample rate.  Rather the group delay is estimated at a lower rate and used to correct for integral tracking errors.

To improve their SNR, the spectrometer phasors $X_i + jY_i$, $i=1\ldots R$, are phase referenced to the white-light phase and coadded as described elsewhere (\cite{PTIvis}).  The group delay is estimated from the peak bin of the complex Fourier transform of the phasors.  Assume the spectrometer is linear in wavenumber from $\nu_1 = 1/\lambda_1$ to $\nu_R = 1/\lambda_R$, for a synthetic wavelength $\Lambda = (\nu_1 - \nu_R)^{-1}$.  With $\lambda_1$ = 2.4~\um\ and $\lambda_2$ = 2.0~\um, $\Lambda$ = 12~\um. Let $f_p$ be the frequency of the peak in the power spectrum
 \begin{equation}
 A(f) = \left | \sum_{i=1}^{R} (X_i+jY_i){\rm e}^{-j2\pi f i} \right | ^2.
 \end{equation} The group delay $g$ is then given as $g = f_p \Lambda (R-1) / R$.

In addition to the group delay itself, a quality measure based on fluctuations in the power spectrum at other frequencies is also estimated.  If the group delay is accepted, a track offset, typically limited to $\pm$1 fringe, is applied to the delay-line target.  Typical coherent integration times for estimating the group delay are 0.4--0.9~s.  To simplify the data processing, track offsets are pended to 0.5~s boundaries.

 \subsection{Other fringe-tracker functions}

The previous discussion addressed the fast-tracking mode of the fringe tracker.  The slow tracking mode will not be discussed in detail here, but a few comments are in order.  With phase referencing and the cold secondary spectrometer, it should be possible to obtain background-limited detection.  This requires a slower readout than described above to minimize the effect of read noise.  At these low rates fringe scanning can be implemented using quarter-wave delay-line offsets rather than with continuous modulation.  Another point is that cophasing quality is not limited to the tracker error, as set by the tracker bandwidth and Greenwood frequency.  Essentially, as the fringe tracker measures its tracking error accurately, that information can be used in an open-loop feedforward correction.  The result is that cophasing quality is limited by time delays from the mean epoch of the fringe measurement to its application to the secondary channel.

Besides tracking and acquisition, the fringe tracker implements initial and on-going calibrations to correct for biases in the fringe parameters as discussed above.  Other functions include alignment modes for positioning the white-light and spectrometer images, and full-frame readout for testing and alignment.  The fringe tracker also implements an FTS mode, which is useful for testing and for wavelength calibration of the spectrometer channels.

 \subsection{Performance}

The photodetection probability for the fringe tracker on PTI is approximately 4\%.  This incorporates all effects, including the factor of 2 attributable to the use of one beamsplitter output for fringe tracking, the 6.75~ms / 10~ms duty cycle of the fringe detector; the spatial filtering attributable to the finite-size white-light pixel; detector quantum efficiency; and warm and cold transmissivity. For $V^2$ = 0.4 and an effective read noise of 12~e$^-$, an SNR for measurement of the fringe phase of 10 and 5 requires 1000 and 400 photons, respectively.  The current fringe-tracking limiting magnitude is 4.5--5.0~K.

 \section{Laser metrology}
 \label{sec:lasermet}
 PTI uses two types of laser metrology systems:  local metrology of the long delay lines for closed-loop servo control, and end-to-end (constant term) metrology for astrometry and control of the short delay line.

 \subsection{Long delay line metrology}

The long delay line metrology system uses a fairly conventional heterodyne system, similar to that used on the Mark~III interferometer (\cite{M3OI}).  The laser source plate uses a stabilized single-frequency HeNe laser and two acousto-optic modulators to generate a 2-MHz frequency offset between polarizations.  The source plate feeds two metrology heads, one each for the active and passive long delay lines.  For each metrology head, one polarization serves as a phase reference while the other polarization measures the delay-line position in double pass.  The metrology light enters and exits through two additional apertures on the delay line front plate.  Flat mirrors, rather than corner reflectors, are used for the metrology endpoints because of their tolerance to beam shear.  A matched pair of optical wedges is used over the metrology apertures on the delay-line front plate so that the metrology light focuses to a slightly different location on the delay-line secondary mirror; thus any laser scatter from secondary-mirror imperfections falls outside of the starlight field-of-view.

The 2-MHz reference signal from the laser source plate and 2-MHz unknown signals from the active and passive long delay line metrology heads are digitized and fed to a custom laser fringe-counter card.  The card counts integral and fractional fringes, providing 32~bits of integer and 8~bits of fraction, for a delay resolution with the double-pass implementation of 1.2~nm over a range of 1300~m.  Limiting doppler excursions to $\pm$1~MHz allows for delay slew rates of $\pm$0.3~m/s.

 \subsection{Constant-term metrology}       
 \label{sec:ct}

Assuming a reasonably stable optical train, the long delay line metrology described above is all that is needed for single-star amplitude observations. Drifts in the unmonitored optical paths, as well as drifts between the (slightly different) metrology and starlight paths through the delay line, are tracked out by the fringe servo and at worst affect only the acquisition time of the fringe tracker.  However, for cophasing and astrometry, metrology of the entire starlight path, common mode with the starlight, is required.  For wide-angle astrometry, both from the ground, and also from space (\cite{SIM}), the interferometer baseline is defined by the location of a corner reflector located on the surface of the siderostat mirror.  External metrology of the corner reflector monitors baseline changes, while internal (constant-term) metrology monitors internal optical paths from the beam combiners to the same point.  However, as discussed in Sec.~\ref{sec:narrowangle}, the requirements on baseline stability for narrow-angle astrometry are much less than for wide-angle astrometry, and it is possible to move the corner cube off the siderostat to a point in the collimated space between the siderostat and DSF.

In the DSF design of Fig.~\ref{fig:dsf}, the SSSM is conjugate to the 40~cm beam from the siderostat at the plane containing the telescope secondary mirror.  This pupil is thus a good location for a retroreflector, as the metrology beam does not shear as the SSSM is adjusted.  Alternately, this plane can be reimaged as shown in the figure.  Lens L1 in Fig.~\ref{fig:dsf} forms an image of this plane (and the SSSM) in collimated space at the location of corner cube CC.  A portion of the metrology light injected from the beam combiners along the primary and secondary paths thus retroreflects off of this common fiducial.

The constant-term metrology system on PTI is a separate heterodyne system from the long delay line metrology system.  The constant-term source plate uses a stabilized laser to generate two dual-polarization beams: one at a frequency difference of 110~kHz at a frequency offset of 27~MHz from the laser frequency for the primary beam combiner, and a second with a frequency difference of 440~kHz at a frequency offset of $-$27~MHz from the laser frequency for the secondary combiner.  The source plate samples each beam to  generate 110 and 440~kHz reference signals.

Initially, consider only the primary constant-term metrology. The 110-kHz dual-frequency beam is injected with a 5-mm mirror into the combined primary starlight beam in the space between beamsplitters B4 and B5, as shown in Fig.~\ref{fig:primary} and retraces the starlight path of each arm to the DSF. Part of the light reflects off the star-separator beamsplitter to corner cube CC, retroreflects, and returns to the primary beam combiner where it is extracted by a dichroic beamsplitter located between beamsplitter B3 and lens L5.  As the starlight beamsplitter is non-polarizing, it is necessary to add polarizers such that the {\it s} polarization traverses one arm and the {\it p} polarization the other.  To achieve this, small polarizers are placed in each arm over the fraction of the pupil used by the metrology beam.  Alternatively, polarizers could be used directly in front of the corner cubes in the DSF (although the polarization state has degraded somewhat by that point).

The extracted metrology beam detected through a cross polarizer is a measure of the difference in the optical pathlengths between the two arms of the interferometer. With the exception of the thin polarizers, this implementation is fully common mode with the starlight, but it is not especially efficient with respect to laser light.  The relatively low polarization difference frequency allows for more sensitive metrology electronics (than those used for the long delay lines) to ensure a good SNR.  The constant-term metrology signal is digitized and measured with a digital fringe counter, similar to the one for the long delay line system.

The secondary constant-term metrology system works identically to the primary one, using the same fiducials. However, examination of Fig.~\ref{fig:dsf} shows that when the SSSM is adjusted so that the primary and secondary beam combiners are boresighted (a calibration mode for dual-star astrometry), the primary metrology light leaks into the secondary path, and vice versa.  The frequency arrangement above, with different difference and offset frequencies for primary and secondary, ensures that the only in-band leakage terms are at the frequency difference of the other metrology system.  These leakage terms are filtered out of the analog signal with bandpass filters.

For differential astrometry the primary and secondary constant-term measurements, ${\rm CT}_p$ and ${\rm CT}_s$, are combined with the phase offsets from the primary and secondary fringe trackers to yield the astrometric delays
 \begin{eqnarray}
 d_p & = & {\rm CT}_p + k^{-1}\Phi_p \\
 d_s & = & {\rm CT}_s + k^{-1}\Phi_s.
 \end{eqnarray}The difference of the two delays, $\Delta = d_p-d_s$, is the narrow-angle astrometric observable.  This difference is also used as the error signal for servo control of the small delay line.  The difference $\Delta$ is calibrated by simultaneous measurements of the same (primary) star by both the primary and secondary fringe trackers.

 \section{Real-time control system}
 \label{sec:rtc}
 The real-time control system on PTI controls the hardware of the individual subsystems as well as orchestrating the interactions between subsystems and providing overall instrument sequencing.   A multiprocessing system is implemented to provide adequate computational resources.  It also allowed a clean partitioning of functionality which enabled concurrent development.

 \subsection{Hardware architecture}

Real-time control and sequencing on PTI is implemented with 7 VME-based 68040 single-board computers (SBCs) interfaced to a SUN workstation, as illustrated in Fig.~\ref{fig:rtc} (\cite{rtc}).  All of the real-time computers use the VxWorks operating system and C/C++. Cardcage 1 contains 4 SBCs, one each for control of the siderostat system, the angle-tracking system, the acquisition system, and overall instrument sequencing.   While the angle-tracker detector interfaces directly to this cardcage using a custom photon-binning board, most of the controlled hardware is located at the siderostat piers.  Small slave card cages at these locations are accessed via fiber-optic bus repeaters to allow low-latency hardware access.  The  remote cardcages contain the GPIB boards for interfacing to the acquisition system, indexers for controlling the siderostats and SSSMs, and D/A converters for driving the FSMs.

The second cardcage contains a single SBC which controls the three servo-controlled delay lines:  the long active, the long passive, and the short active (the short passive delay line has no actuators).  This system implements a delay-line controller, described elsewhere (\cite{dl1}; \cite{dl2}).

The third cardcage contains two SBCs, one each for the primary and secondary fringe trackers.  The cardcage contains custom boards for clocking and interface to the IR arrays, as well as video cards for array display.

Inter-cardcage communication is via a reflective memory (essentially an asynchronous shared memory).  The software driver uses the reflective memory to implement FIFOs, double-buffers, and other structures for interprocess communication.  Absolute time is available in each cardcage via an IRIG-B time-code reader; IRIG-B time comes from a GPS system. While each SBC uses ethernet for booting, only the sequencer implements a real-time ethernet connection to the SUN; all other subsystems interface to the sequencer (and each other) via the reflective memory.

 \subsection{Software architecture}

The software architecture for each subsystem is similar, consisting of a number of periodically-scheduled tasks (1--2000 Hz) tied to GPS time, as well as event-driven tasks. Each subsystem implements a state machine characterized by modes which can be commanded, e.g., for the siderostat subsystem: {\it track}, {\it home}, etc., and associated states sequenced by the individual subsystem, e.g. {\it track:slew}, {\it track:track}, etc. Each subsystem provides a library of public functions that can be called by other subsystems. For the siderostat example, public functions are available to pass pointing handoffs, provide tracking corrections, etc. These public functions use the interprocessor-communications layer to call private functions that provide the actual functionality. Common public functions for each subsystem support command interpretation, data recording, and status and error reporting.

The sequencer is the only subsystem without explicit hardware. It provides high-level instrument sequencing and apparent-place calculations,  automating observations to provide high observational throughput. The sequencer communicates with the hardware subsystems through their public functions, and implements a command relayer and  data, status, and error servers. The sequencer interfaces with the SUN host using the RPC protocol over ethernet.  RPCs are used to implement a command stream from the SUN to the sequencer, and to return error, status, and data streams. The SUN hosts a graphical user interface for instrument control and monitoring, as well as the data-recording system.

 \subsection{User interface and sequencing}

The user interface provides several views into the system.  The highest-level view is the observing window, which provides a summary of the status of each subsystem and control of the overall instrument sequencing.  The instrument sequencer, a large state machine running on the sequencer SBC, provides fully automated observing (after nightly setup) based on star lists describing the targets and integration times.  User control at this level is of the overall sequencer and data recorder. The sequencer is multithreaded, allowing concurrency to speed operations.  For example, for automating a single-star observation, concurrent threads implement delay-line slew, and star acquisition and tracking; when the threads rendezvous, fringe acquisition can begin.

For engineering and debugging, each subsystem provides a detailed window which displays a variety of status information and provides direct control of the subsystem mode, subsystem data recording, and access to many public functions, for example, to allow the passing of manual handoffs.  All systems have a parameter structure that contains hardware mappings, servo parameters, etc., which can be updated from the engineering window.  The default values of the parameter structures are updated from a free-form initialization file.

The data-recording subsystem separates the combined data-recording stream, tags the data blocks, and generates separate files for each subsystem or instance of a subsystem.  These files are recorded to disk and later transferred to CD-ROM for archiving.

 \acknowledgments

Funding for the development of PTI was provided by NASA under its TOPS (Toward Other Planetary Systems) and ASEPS (Astronomical Studies of Extrasolar Planetary Systems) programs, and from the JPL Director's Discretionary Fund.  Ongoing funding has been provided by NASA through its Origins Program and from the JPL Directors Research and Development Fund.  Thanks to Kadri Vural and Rockwell International for the NICMOS-3 detectors.  Thanks also to the contributions of Chas Beichman, Gary Brack, Carl Bruce, Rob Calvet, Jean-Francois Leger, Harjit Singh, Sylvain Takerkart, Fred Vescelus, and Randy Wager.  The work performed here was conducted at the Jet Propulsion Laboratory, California Institute of Technology, under contract with the National Aeronautics and Space Administration.

 \appendix

 \section{Coherence time}
 \label{sec:phase}

Coherence time can be defined in various ways (\cite{Buscher}).  Let $\tau_{0,i}$ denote the structure-function definition of coherence time, viz.\ that sample spacing for which the phase difference between samples is one radian rms.  The structure function depends on time as $D_i(t) = (t/\tau_{0,i})^{5/3}$.  For $i=1$, representing contributions from a single point on the wavefront (the usual adaptive-optics definition), $\tau_{0,1}$ = $0.314r_0/W$ for coherence diameter $r_0$ and constant wind speed $W$ (\cite{Buscher}).  For $i=2$, applicable to interferometry, there are contributions from two apertures, and $\tau_{0,2} = 0.207r_0/W$.  Both definitions assume an outer scale much larger than $\tau_{0,i} W$, generally a good approximation.  The definition of $\tau_{0,2}$ assumes an interferometer baseline $B \gg \tau_{0,2}W$, valid for the long baselines of PTI.

Let $T_{0,i}$ to denote the variance definition of coherence time, viz.\ that time interval for which the phase fluctuations about the interval mean are one radian rms.  It is given by $T_{0,1} = 1.235r_0/W$ and $T_{0,2} = 0.815r_0/W$, with time evolution $\sigma_i^2=(T/T_{0,i})^{5/3}$.  One- and two-aperture Greenwood frequencies (\cite{Greenwood}) can be computed similarly as $f_{G,1}=(2.343r_0/W)^{-1}$ and $f_{G,2} = (1.546r_0/W)^{-1}$.  For a single-pole servo with bandwidth $f_c$, the servo error variance is given by $\epsilon_i^2 = (f_{G,i}/f_c)^{5/3}$ rad$^2$.

 \section{Estimating the fringe phase}
 \label{sec:kf}

We model the time evolution of the total fringe phase $\chi(t)$ by its structure function $D_2(t) = (t/\tau_{0,2})^{5/3}$, valid for $t \ll B/W$, where $B$ is the interferometer baseline.  The corresponding correlation function $C(t)$ is given by (see \cite{Ishimaru}) $C_2(t) = C_2(0) - 0.5D_2(t)$.  A state-space representation of the fringe phase process can be formed using Levinson's recursion to generate the autoregressive coefficients (see \cite{KayMarple}). In the limit of long baselines, $C_2(0) \rightarrow \infty$, and the one-state model for a sampling spacing $t$ is given by
 \begin{eqnarray}
 x_{n+1} & = & x_n + w_n \\
 \chi_n  & = & x_n,
 \end{eqnarray}
 with plant noise $\sigma_w^2 = D_2(t)$.  For noiseless observations, the best one-step predictor of the fringe phase is $\widehat{\chi}_{n+1|n} = \chi_{n}$, i.e., just the previous point, with a prediction error given by the plant noise $D_2(t)$.

A two-state model is given by
 \begin{eqnarray}
 x_{n+1} & = & \left[ \begin{array}{cc} 0 & 1 \\ -0.587 & 1.587 \end{array} \right] x_n
 + \left [\begin{array}{c}  0 \\ 1 \end{array} \right] w_n \\
 \chi_n & = & \left[ \begin{array}{cc} 0 & 1 \end{array} \right] x_n,
 \end{eqnarray}
 with plant noise $\sigma_w^2 = 0.655D_2(t)$.  For noiseless observations, the best one-step predictor is $\widehat{\chi}_{n+1|n} = \chi_n + 0.587(\chi_n-\chi_{n-1})$, with a prediction error given by the plant noise $0.655D_2(t)$. Thus including a velocity term substantially reduces the prediction error. Incorporating the factor of 0.655 into the structure function, we can write the plant noise (and prediction error) as $\sigma_w^2 = D^{\prime}_2(t)$, where $D^{\prime}_2(t) = (t/\tau_{0,2}^{\prime})^{5/3}$ with an effective coherence time $\tau_{0,2}^{\prime} = 1.289\tau_{0,2}$.

These matrix forms allow for a straightforward Kalman-filter implementation (see \cite{Gelb}), which can accommodate observation noise in measuring the fringe.  Higher-order models provide little additional benefit: a three-state model has a prediction variance only 2\% better than the two-state model.

 \section{Atmospheric dispersion}
 \label{sec:disp}

Use of an air delay line can introduce visibility losses, as the (dispersionless) vacuum delay of an off-zenith source is compensated by a dispersive air path. Fortunately, atmospheric dispersion is relatively small in the infrared.  The second column of Table~\ref{tab:disp} gives the residual group delay as a function of wavelength when the phase delays are matched at 2.2~\um.  These calculations assume a maximum air delay of 40~m at 0.8~atm.  At $\lambda$ = 2.2~\um, the fringe envelope is offset 16~\um\ from the point of equal phase delays between the air and vacuum paths, with a spread of 12~\um\ between 2.0 and 2.4~\um.  Matching group delays between the air and vacuum paths at 2.2~\um\ yields the third column of Table~\ref{tab:disp}, which shows the residual phase as a function of wavelength about this point.  Over the K~band (2.0--2.4~\um), the quadratic phase residual yields a $\Delta V$ = 1.5\% reduction in broadband (white-light) fringe visibility.

For the narrow spectrometer channels, the spread in group delay across the band introduces only small visibility losses.  For a spectrometer resolution $R$, the visibility loss attributable to an offset $x$ from the peak of the fringe envelope for a particular channel is $\Delta V \simeq (\pi x /(\lambda R))^2 / 6$; with a 5~\um\ offset and $R$ = 40, the visibility reduction is $\sim$0.5\%.

 \section{Fringe demodulation with fixed-width time bins}
 \label{sec:fixed}

When the effective pathlength-modulation stroke is not exactly equal to the wavelength, the fringe phasor estimates are distorted.  However, they can be corrected in a straightforward manner.

Write the fringe pattern as
 \begin{equation}
 f(\phi) = \frac{1}{s}\left[X^*\cos{\phi}+Y^*\sin{\phi}+N^*\right],
 \end{equation}
 where $s$ is the length of the pathlength modulation stroke, ${X^*}^2+{Y^*}^2 =  (N^*V)^2$, and $\phi=\tan^{-1}(Y^*/X^*)-\pi/2$ ($\sin$ fringe). Define four time bins $A=[-s/2,-s/4]$, $B=[-s/4,0]$, $C=[0,s/4]$, and $D = [s/4,s/2]$, and let  $X = A-C$, $Y = B-D$, $N = A+B+C+D$. Then 
 \begin{eqnarray}
 X & = & (1/s)[-\alpha(s)X^* - \beta(s)Y^*] \\
 Y & = & (1/s)[+\alpha(s)X^* - \beta(s)Y^*] \\
 N & = & (1/s)[+\gamma(s)X^* + sN^*],
 \label{eq:xyn}
 \end{eqnarray}
 where
 \begin{eqnarray}
 \alpha(s) & = & 2\sin(s/4) - \sin(s/2) \\
 \beta(s)  & = & 1-\cos(s/2) \\
 \gamma(s) & = & 2\sin(s/2).
 \end{eqnarray}
 For a matched stroke ($s=2\pi$), $\alpha = \beta = 2.$

The measured values $X$, $Y$, and $N$, can be corrected for the mismatched stroke as
 \begin{eqnarray}
 X_{\rm corr} & = & \frac{X + Y}{\beta} + \frac{X - Y}{\alpha} \\
 Y_{\rm corr} & = & \frac{X + Y}{\beta} - \frac{X - Y}{\alpha} \\
 N_{\rm corr} & = & N + \gamma^{\prime}(X - Y),
 \end{eqnarray}
 where $\gamma^{\prime}$ = $\gamma/(2\alpha)$.  While $X_{\rm corr}$, $Y_{\rm corr}$, and $N_{\rm corr}$ are unbiased, their variances are more complicated than for the case of a matched stroke, and there is generally some performance penalty.

 \clearpage



\clearpage

 \figcaption[]{Photograph of the Palomar Testbed Interferometer (PTI) at Palomar Observatory taken from the catwalk of the 5-m Hale telescope looking northeast.  The driveway and adjacent road are painted white to minimize daytime heating. \label{fig:photo}}

\figcaption[]{Differential astrometric accuracy vs.\ star separation in a one-hour integration for different baseline lengths (from \protect\cite{AA92}); atmospheric models providing 1/2- and 1.0-arcsec seeing are shown.  These results assume an infinite outer scale, and better results are achieved when the baseline exceeds the outer scale, as would be expected with a 100-m baseline at most sites.  Measurements with the Mark~III interferometer of a 3.3-arcsec binary star (\protect\cite{AA94}) are consistent with the model. \label{fig:atmos}}

 \figcaption[]{Dual-star concept. \label{fig:dualstar}}

 \figcaption[]{PTI beam train.  \label{fig:beamtrain}}

 \figcaption[]{PTI building and overall layout. \label{fig:building}}

 \figcaption[]{Schematic of the acquisition system. \label{fig:acq}}

 \figcaption[]{Schematic of the dual-star feed.  Mirror M2 is the field separator; mirrors M4 and M4\p\ are also fast-steering mirrors; mirror M6\p\ is the secondary star selector mirror (SSSM). \label{fig:dsf}}

 \figcaption[]{Detail of the beam-combining lab. Light paths for primary (solid line) and secondary (dotted line) stars are shown.  The south delay line feed mirrors can be removed to allow injection of light from the west siderostat.  The beams exiting the long delay lines pass beneath the delay line feed mirrors into the beam compressors.   \label{fig:detail}}

\figcaption[]{Sky coverage for PTI.  The dashed lines bound the sky coverage for the N-S baseline as limited by the delay line range; the solid lines bound the sky coverage for the N-W baseline.  The shaded areas on the figure indicate where areal vignetting of the beam exceeds 20\%.  \label{fig:sky}}

 \figcaption[]{Schematic of the primary beam-combiner table.  \label{fig:primary}}

 \figcaption[]{Layout of the secondary dewar. \label{fig:secondarydewar}}

 \figcaption[]{PTI computer hardware architecture.  \label{fig:rtc}}

 \clearpage

 \begin{table*}
 \caption{PTI vital statistics}
 \label{tab:PTIstats}
 \begin{center}
 \begin{tabular}{ll}
 \\
 Location	& Palomar Observatory, Palomar Mountain, CA \\		& (long: -116.8633\arcdeg; lat: 33.3567\arcdeg; elevation: 1687 m) \\
 Baselines	& NS: 110~m: ( 37.1 E, 103.3 N, -3.3 Z) \\& NW:  86~m: (-81.7 E, -28.2 N,  3.1 Z) \\& SW:  87~m: (-44.6 E,  75.1 N, -0.2 Z) \\
 Aperture	& 40 cm \\
 Delay range	& $\pm$ 38.3 m \\
 Architecture	& dual-star, active fringe tracking \\
 Fringe-tracking wavelength 	& 2.0--2.4 \um \\
 Spectrometer wavelength 	& 1.5--2.4 \um \\
 Fringe sensor & NICMOS-3 infrared array with fringe-scanning \\
 		   & modulation, 10~ms frame rate \\
 Angle-tracking wavelength & 0.7--1.0 \um \\
 Angle sensor	& Si APD quad cell \\
 \end{tabular}
 \end{center}
 \end{table*}

 \clearpage

 \begin{table*}
 \caption{Residual group and phase delays attributable to atmospheric dispersion at maximum delay-line position (see App.~\ref{sec:disp}).}
 \label{tab:disp}
 \begin{center}
 \begin{tabular}{ccc}
 \\
 $\lambda$ (\um) & residual group delay (\um) & residual phase (rads) \\
     1.6       & 38.04     & 11.05 \\
     1.7       & 32.74     & 6.63 \\
     1.8       & 28.31     & 3.70 \\
     1.9       & 24.56     & 1.83 \\
     2.0       & 21.37     & 0.72 \\
     2.1       & 18.62     & 0.16 \\
     2.2       & 16.24     & 0.00 \\
     2.3       & 14.16     & 0.13 \\
     2.4       & 12.34     & 0.47 
 \end{tabular}
 \end{center}
 \end{table*}

\clearpage

\clearpage
\begin{figure}
\plotone{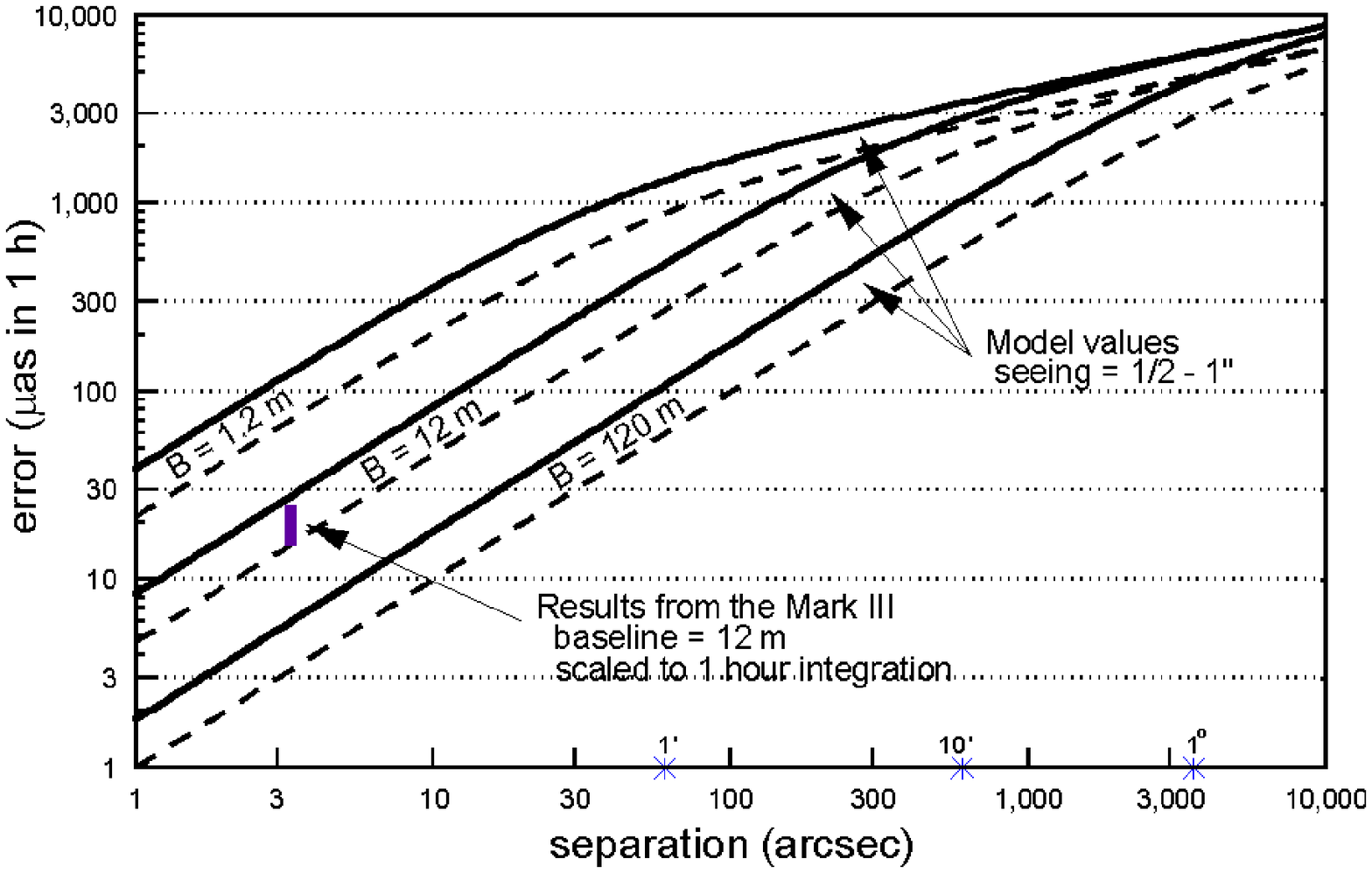}
\end{figure}

\clearpage
\begin{figure}
\plotone{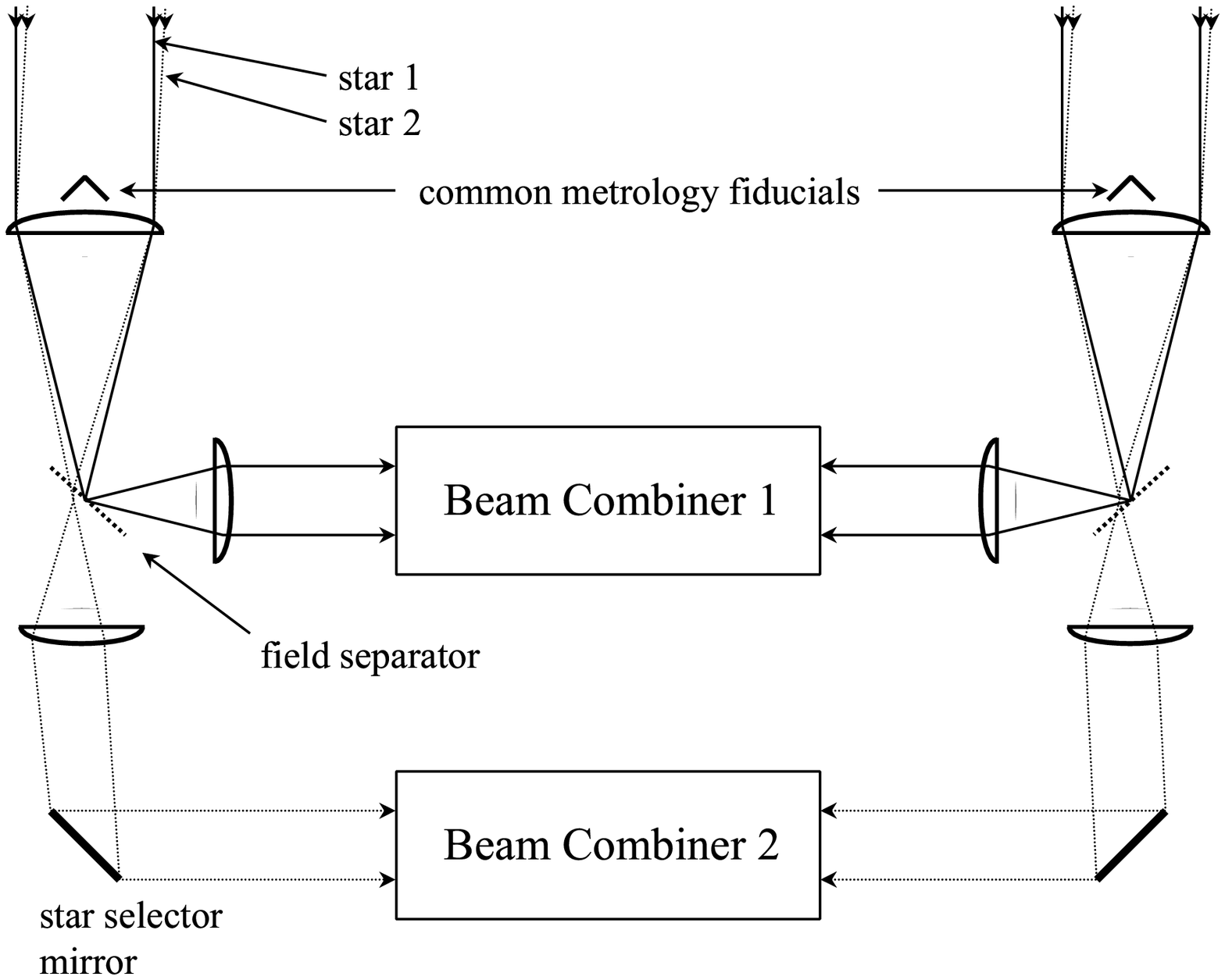}
\end{figure}

\clearpage
\begin{figure}
\plotone{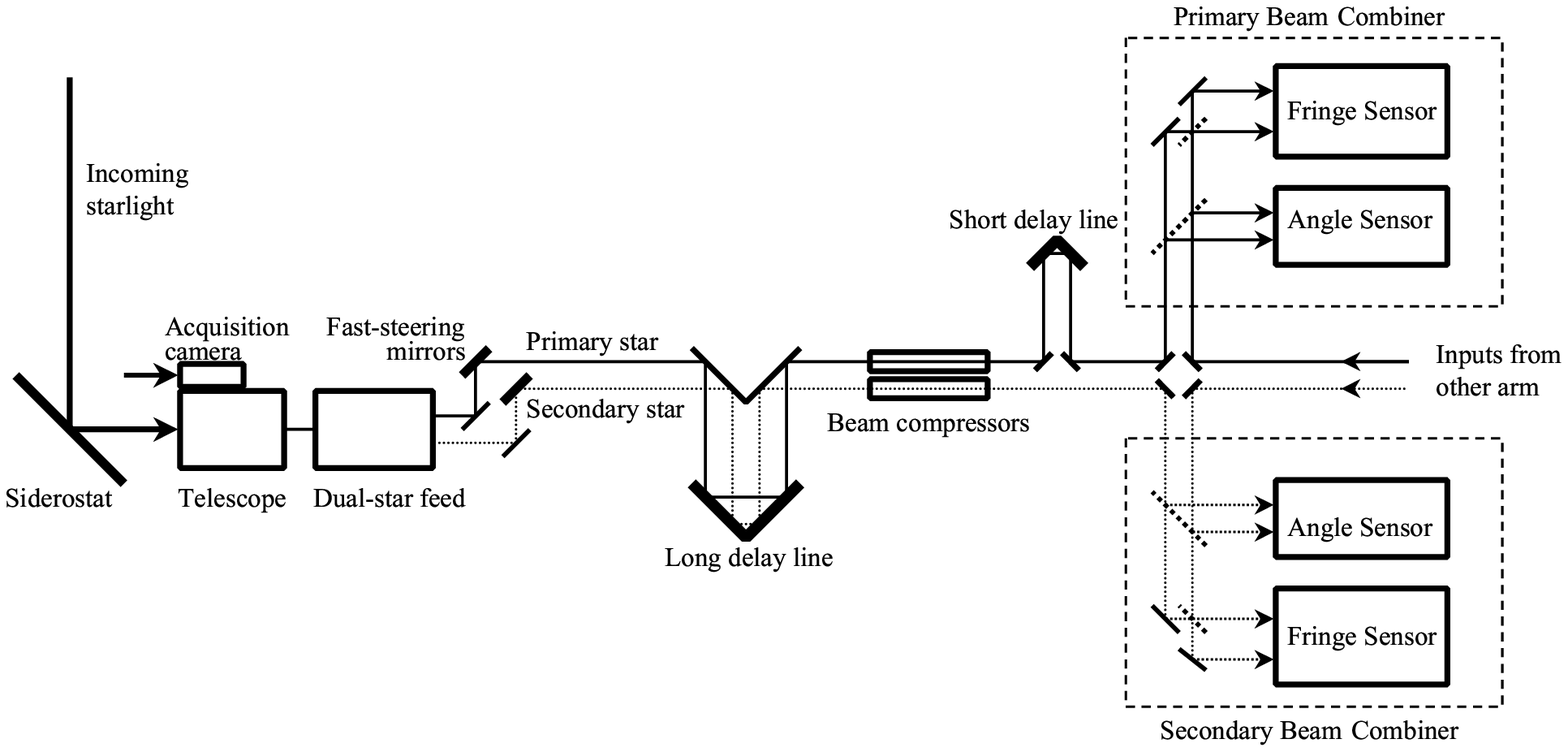}
\end{figure}

\clearpage
\begin{figure}
\plotone{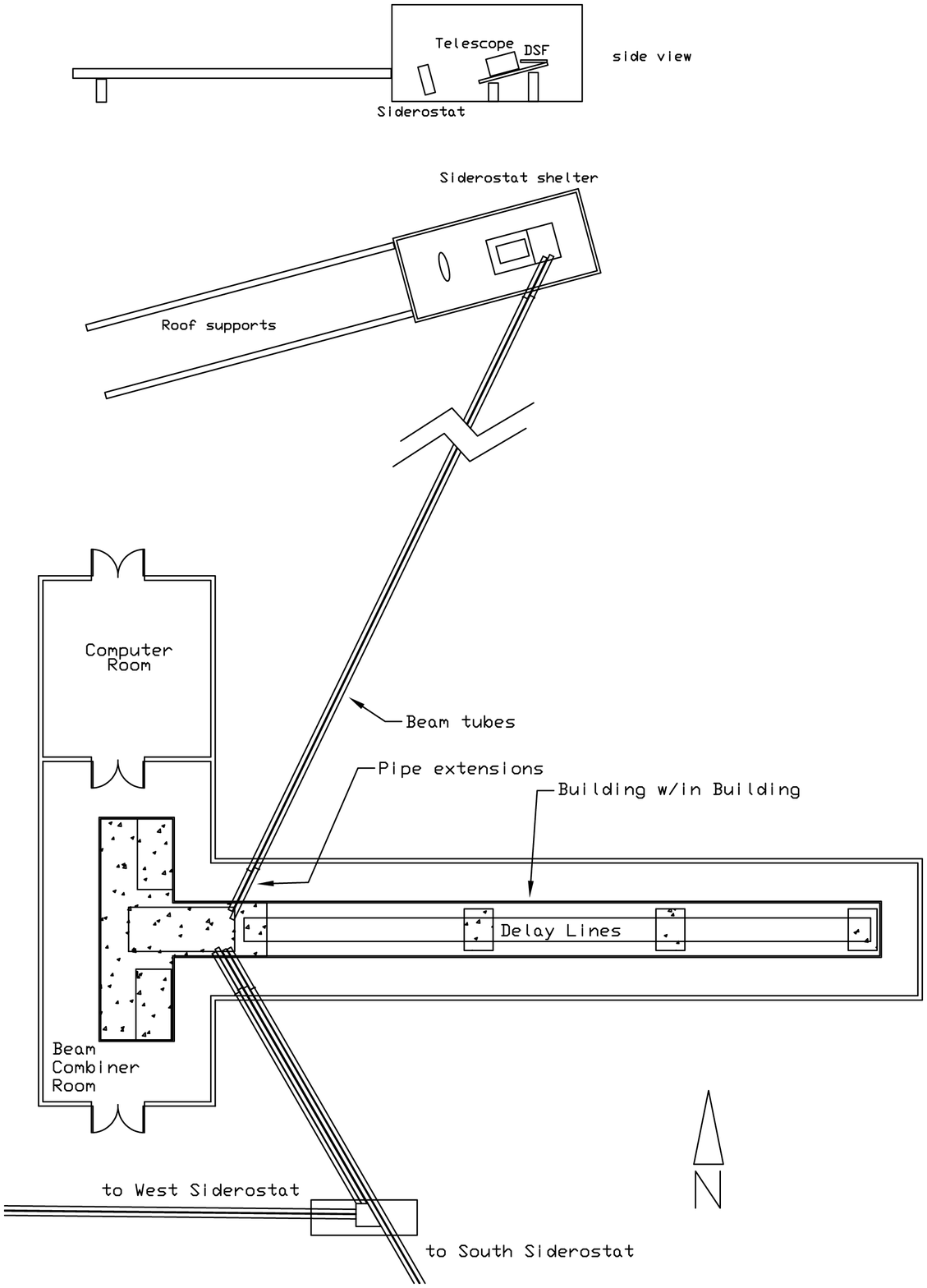}
\end{figure}

\clearpage
\begin{figure}
\plotone{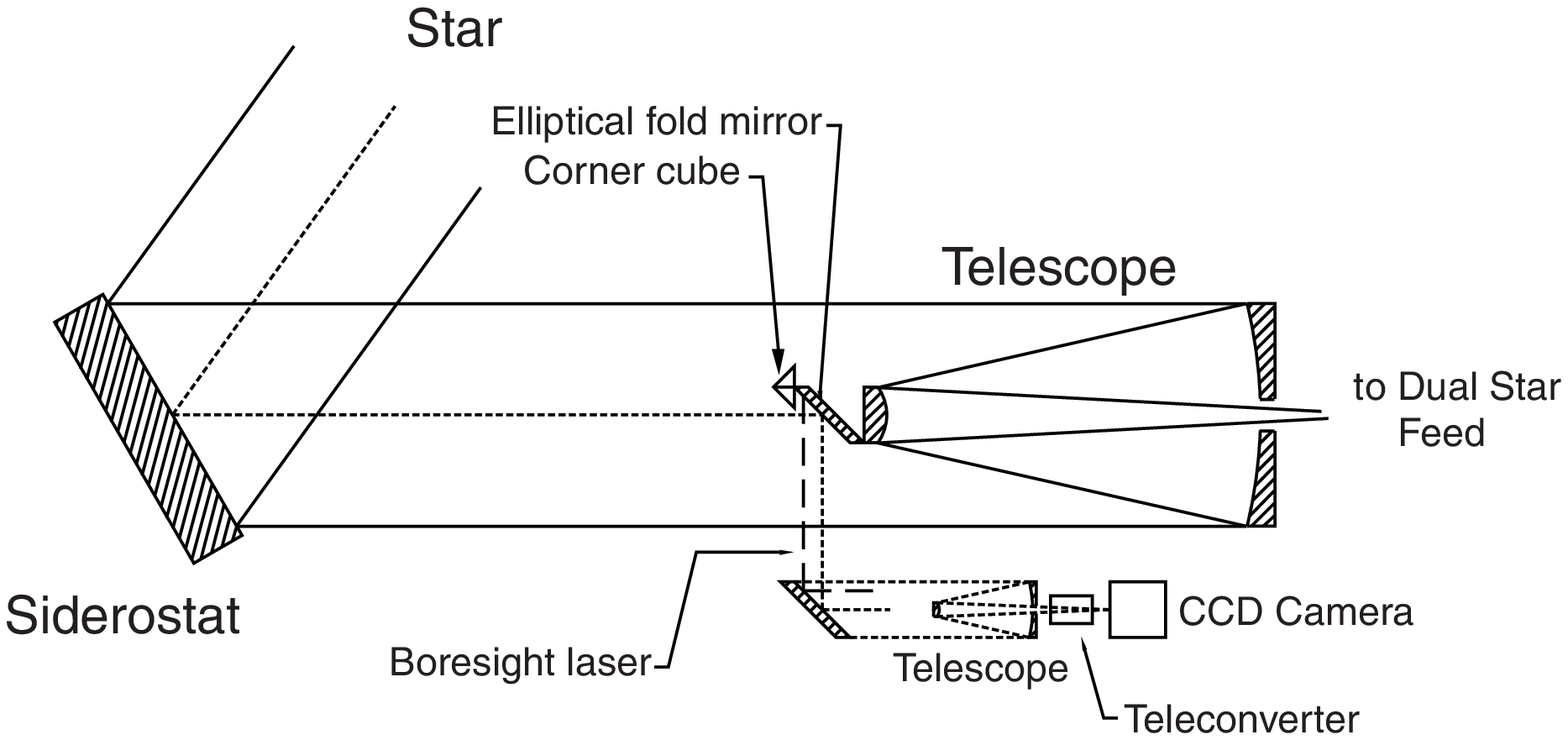}
\end{figure}

\clearpage
\begin{figure}
\plotone{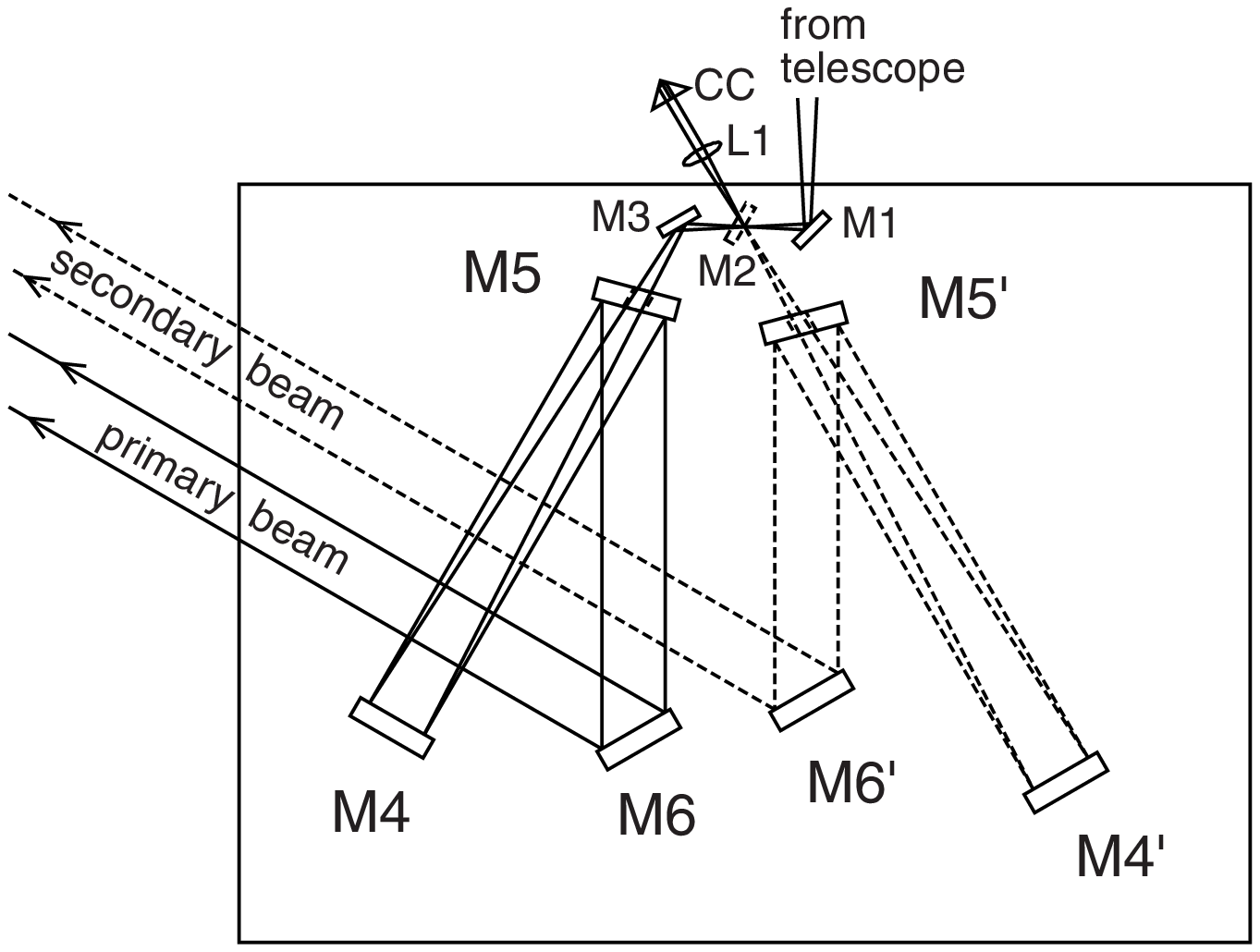}
\end{figure}

\clearpage
\begin{figure}
\plotone{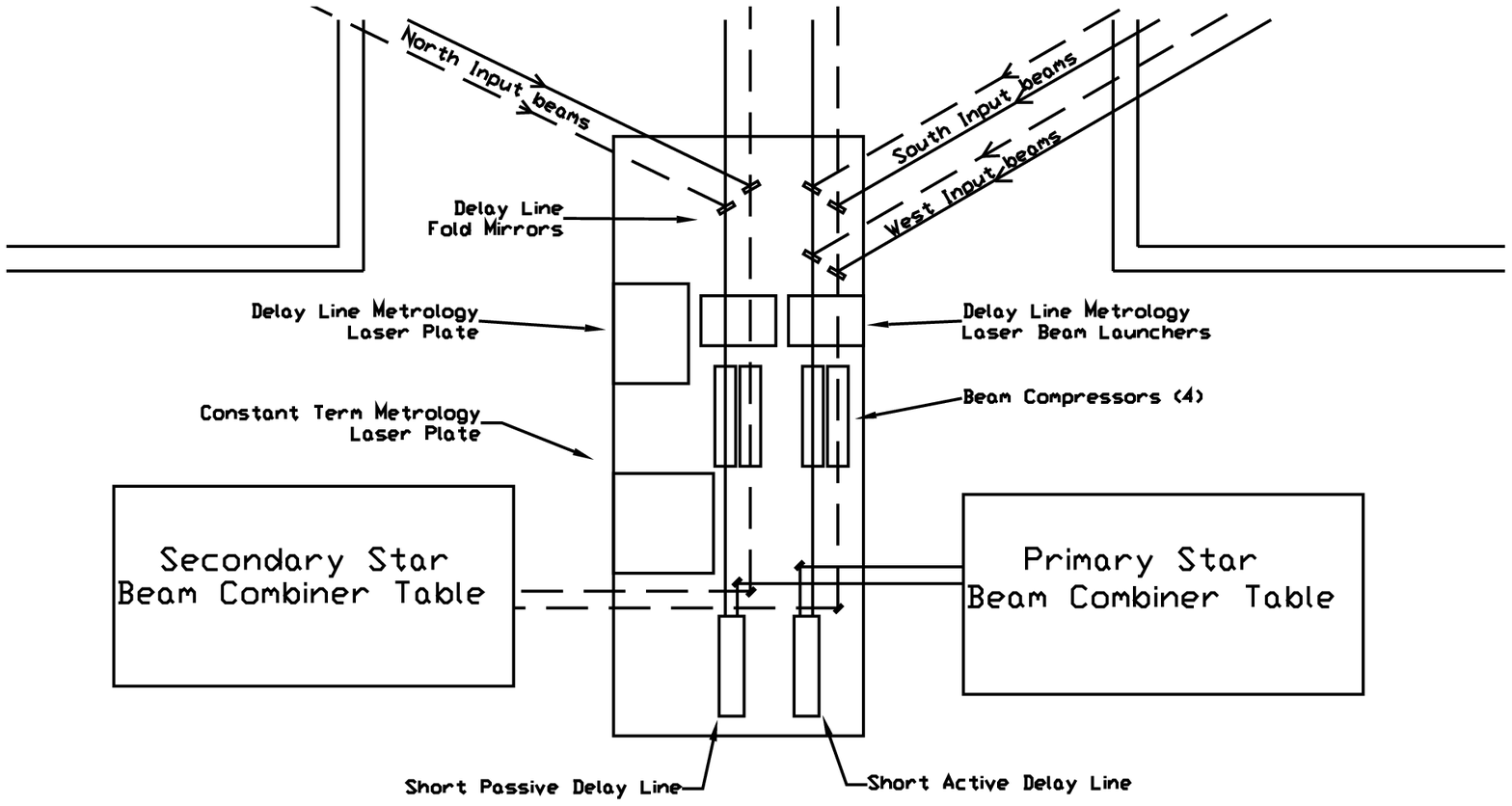}
\end{figure}

\clearpage
\begin{figure}
\plotone{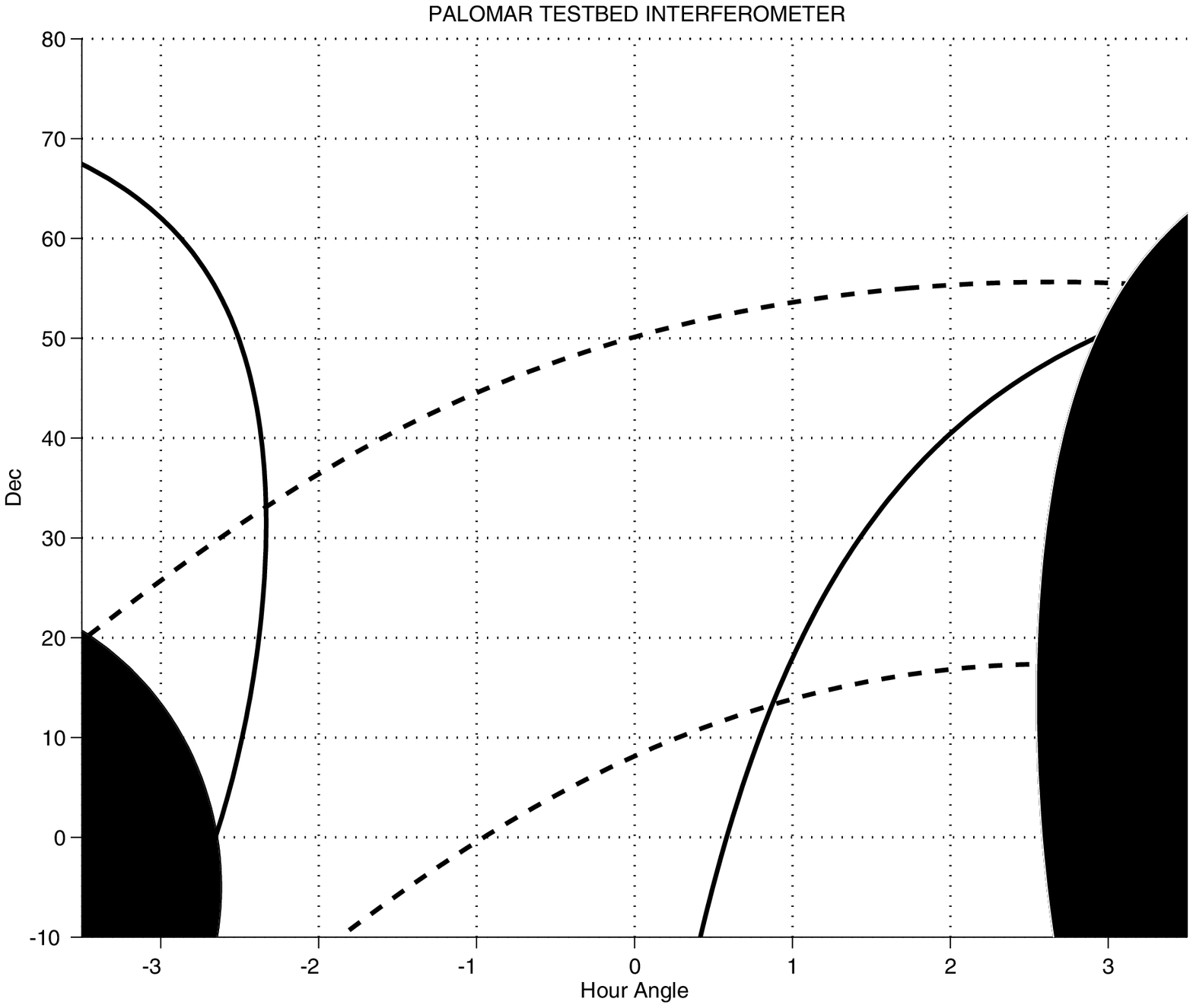}
\end{figure}

\clearpage
\begin{figure}
\plotone{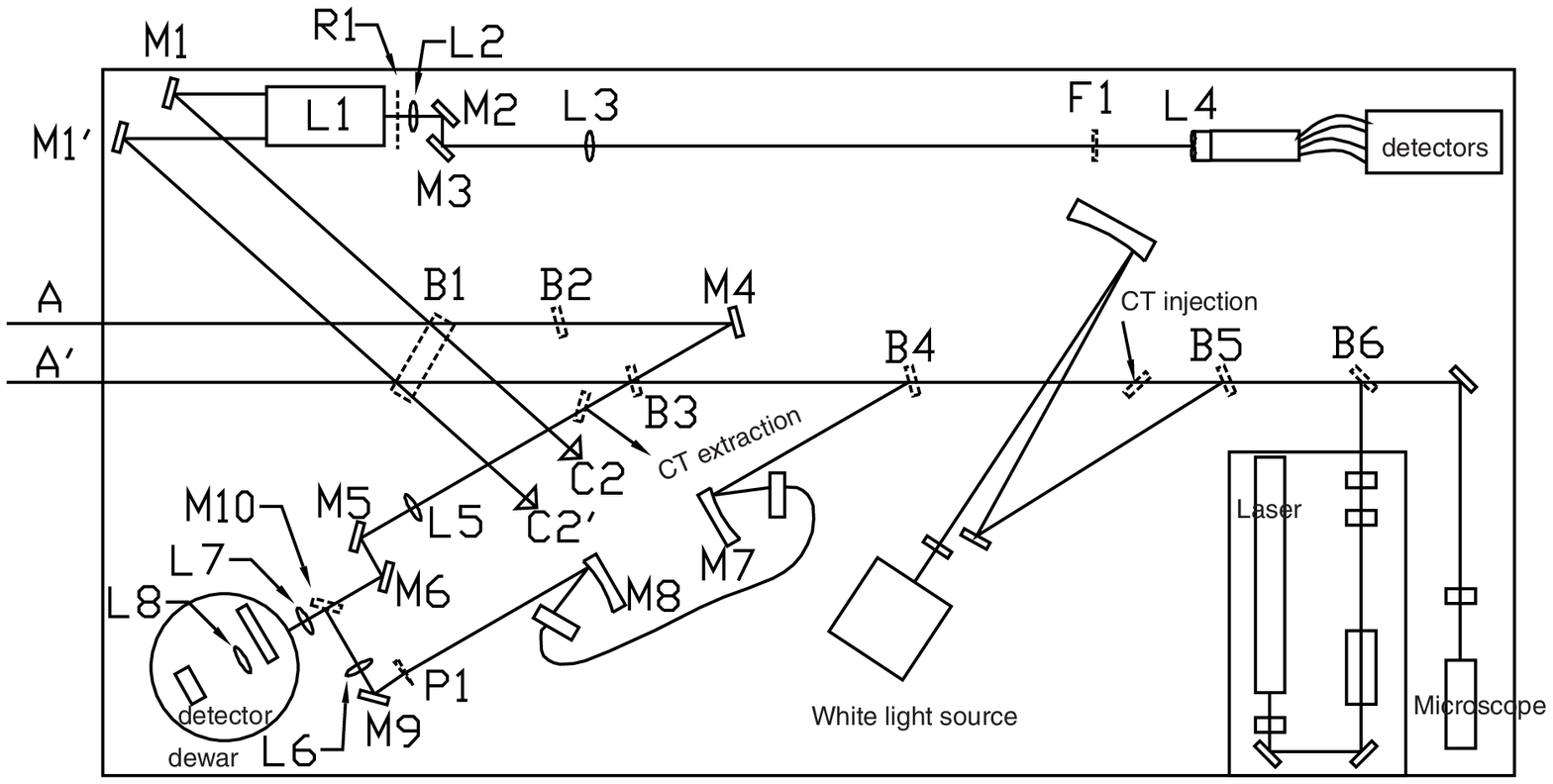}
\end{figure}

\clearpage
\begin{figure}
\plotone{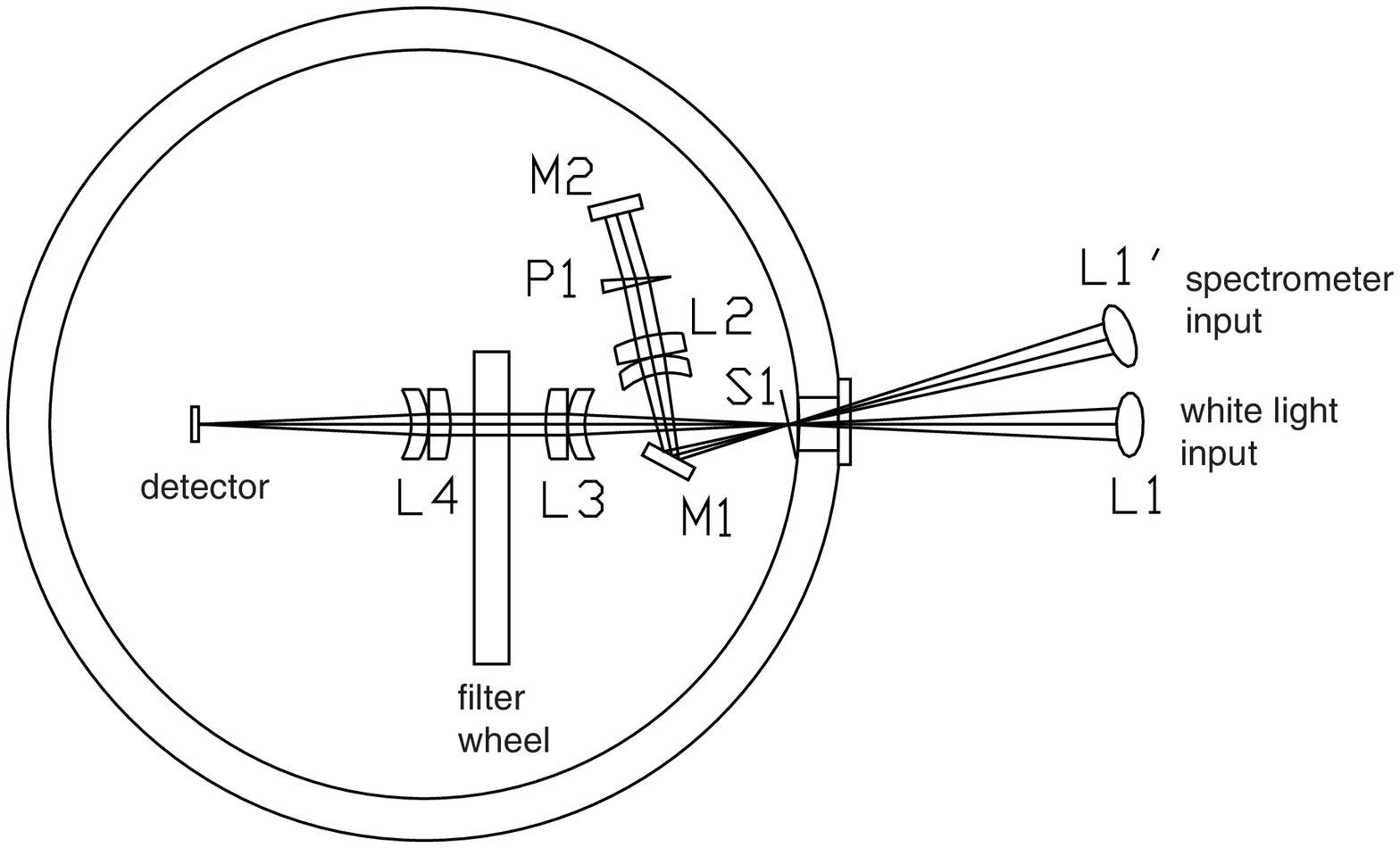}
\end{figure}

\clearpage
\begin{figure}
\plotone{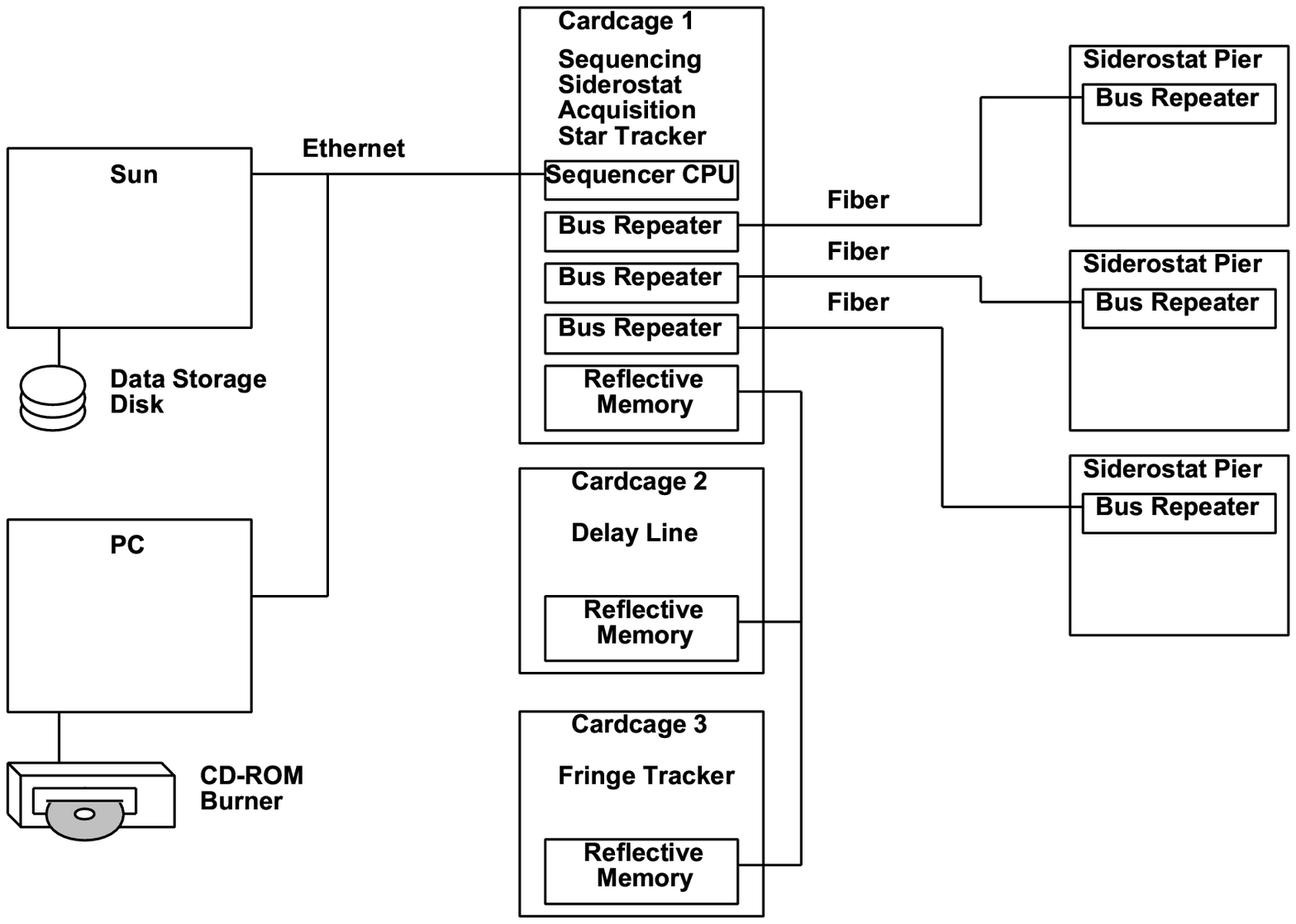}
\end{figure}

\clearpage

\end{document}